\documentclass[structabstract]{aa}

\usepackage{natbib}
\usepackage{graphicx}
\usepackage{amsmath}
\usepackage{amsbsy}
\usepackage{txfonts}

\newcommand{\micron}{$\mu$m}

\newcommand{\ten}[1]{$10^{#1}$}
\newcommand{\scit}[2]{$#1\times10^{#2}$}
\newcommand{\scim}[2]{#1\times10^{#2}}

\newcommand{\ps}{s$^{-1}$}

\newcommand{\pcs}{cm$^{-2}$}
\newcommand{\pcc}{cm$^{-3}$}

\newcommand{\kkms}{K km s$^{-1}$}

\newcommand{\tmbdv}{\int T_{\rm mb}{\rm d}\varv}
\newcommand{\fnudnu}{\int F_\nu{\rm d}\nu}
\newcommand{\eq}[1]{Eq.\ (\ref{eq:#1})}

\newcommand{\fig}[1]{Fig.\ \ref{fig:#1}}
\newcommand{\figg}[1]{Figure \ref{fig:#1}}
\newcommand{\tb}[1]{Table \ref{tb:#1}}

\newcommand{\cellcent}[1]{\multicolumn{1}{c}{#1}}
\newcommand{\pz}{\phantom{0}}

\newcommand{\q}[1]{_{\rm #1}}
\newcommand{\av}{A_{\rm V}}
\newcommand{\tauuv}{\tau_{\rm UV}}
\newcommand{\jup}{J_{\rm u}}
\newcommand{\eup}{E_{\rm u}}
\newcommand{\mh}{H$_2$}
\newcommand{\mhm}{{\rm H}_2}
\newcommand{\w}{H$_2$O}
\newcommand{\mn}{N$_2$}

\newcommand{\chm}[1]{\boldsymbol{#1}}

\defcitealias{kaufman99a}{K99}
\newcommand{\knn}{\citetalias{kaufman99a}}

\begin{document}

\title{Modelling \emph{Herschel} observations of hot molecular gas emission from embedded low-mass protostars\thanks{\emph{Herschel} is an ESA space observatory with science instruments provided by European-led Principal Investigator consortia and with important participation from NASA.}}

\author{
R. Visser\inst{1,2}
\and L.E. Kristensen\inst{1}
\and S. Bruderer\inst{3,4}
\and E.F. van Dishoeck\inst{1,3}
\and G.J. Herczeg\inst{3}
\and C. Brinch\inst{1}
\and S.D. Doty\inst{5}
\and D. Harsono\inst{1}
\and M.G. Wolfire\inst{6}
}

\institute{
Leiden Observatory, Leiden University, P.O.\ Box 9513, 2300 RA Leiden, the Netherlands
\and
Department of Astronomy, University of Michigan, 500 Church Street, Ann Arbor, MI 48109-1042, USA\\
  \email{visserr@umich.edu}
\and
Max-Planck-Institut f\"ur extraterrestrische Physik, Giessenbachstrasse 1, 85748 Garching, Germany
\and
Institute of Astronomy, ETH Zurich, Wolfgang-Pauli-Strasse 27, 8093 Zurich, Switzerland
\and
Department of Physics and Astronomy, Denison University, Granville, OH 43023, USA
\and
Department of Astronomy, University of Maryland, College Park, MD 20742-2421, USA
}

\titlerunning{Modelling \emph{Herschel} observations of hot molecular gas from embedded protostars}

\date{Accepted version \today}


\abstract
{} 
{Young stars interact vigorously with their surroundings, as evident from the highly rotationally excited CO (up to $\eup/k=4000$ K) and \w{} emission (up to 600 K) detected by the \emph{Herschel Space Observatory} in embedded low-mass protostars. Our aim is to construct a model that reproduces the observations quantitatively, to investigate the origin of the emission, and to use the lines as probes of the various heating mechanisms.} 
{The model consists of a spherical envelope with a power-law density structure and a bipolar outflow cavity. Three heating mechanisms are considered: passive heating by the protostellar luminosity, ultraviolet irradiation of the outflow cavity walls, and small-scale C-type shocks along the cavity walls. Most of the model parameters are constrained from independent observations; the two remaining free parameters considered here are the protostellar UV luminosity and the shock velocity. Line fluxes are calculated for CO and \w{} and compared to \emph{Herschel} data and complementary ground-based data for the protostars NGC1333 IRAS2A, HH~46 and DK~Cha. The three sources are selected to span a range of evolutionary phases (early Stage 0 to late Stage I) and physical characteristics such as luminosity and envelope mass.} 
{The bulk of the gas in the envelope, heated by the protostellar luminosity, accounts for 3--10\% of the CO luminosity summed over all rotational lines up to $J=40$--39; it is best probed by low-$J$ CO isotopologue lines such as C$^{18}$O 2--1 and 3--2. The UV-heated gas and the C-type shocks, probed by $^{12}$CO 10--9 and higher-$J$ lines, contribute 20--80\% each. The model fits show a tentative evolutionary trend: the CO emission is dominated by shocks in the youngest source and by UV-heated gas in the oldest one. This trend is mainly driven by the lower envelope density in more evolved sources. The total \w{} line luminosity in all cases is dominated by shocks ($>99\%$). The exact percentages for both species are uncertain by at least a factor of 2 due to uncertainties in the gas temperature as function of the incident UV flux. However, on a qualitative level and within the context of our model, both UV-heated gas and C-type shocks are needed to reproduce the emission in far-infrared rotational lines of CO and \w.} 
{} 


\maketitle


\section{Introduction}
\label{sec:intro}
In the embedded phases of low-mass star formation 
\citep[Stages 0 and I;][]{whitney03b,robitaille06a},
the material surrounding the protostar is exposed to energetic phenomena such as shocks and ultraviolet radiation \citep{shu87a,spaans95a,bachiller99a,arce07a}.
Observationally, this leads to much higher intensities in lines of rotationally excited carbon monoxide, water and other species than would be possible from gas that is only heated through collisions with warm dust in the bulk of the envelope. The ubiquity of surplus rotationally excited emission was first established through ground-based observations of the CO $J=6$--5 transition \citep[upper-level energy $\eup/k=120$ K;][]{schuster93a,hogerheijde98a}. The \emph{Infrared Space Observatory} (ISO) subsequently detected CO up to the 21--20 line ($\eup/k=1300$ K), along with rotationally excited \w{} (up to 470 K) and OH \citep[up to 620 K;][]{ceccarelli99a,giannini99a,giannini01a,nisini02a}. The origin of this hot gas has been debated ever since. The shape of the CO 6--5 spectra -- a narrow emission feature on top of a broader one -- and the presence of a strong, narrow $^{13}$CO 6--5 emission feature point at a mixture of quiescent and shocked gas, with the quiescent gas likely heated by the protostellar UV field along the cavity walls \citep{spaans95a}. Spatially resolved CO 6--5 maps, showing extended narrow emission, provide additional support for this scenario \citep{vankempen09b}. However, it remains unclear at a quantitative level to what extent UV radiation and shocks each contribute to heating up the gas and powering the emission.

The \emph{Herschel Space Observatory} \citep{pilbratt10a} offers a new set of tools to tackle this question. Amongst its first results are detections of highly rotationally excited CO (up to $J=38$--37, $\eup/k=4100$ K) and \w{} (up to 640 K) in the embedded low-mass protostars \object{HH~46} and \object{DK~Cha} \citep{vankempen10b,vankempen10a}. A quantitative radiative transfer model of HH~46 was successful at explaining the CO line fluxes with a combination of UV-heated gas in the outflow cavity walls and small-scale C-type shocks along the walls \citep{vankempen10a}. However, the model was unable to disentangle the origin of the observed \w{} lines in a similar fashion. The goals of the current paper are (1) to present our model from \citeauthor{vankempen10a} in more detail; (2) to check how robust the model fit is for CO; (3) to derive a quantitative fit for \w; and (4) to extend the analysis to DK~Cha and a third source (\object{NGC1333 IRAS2A}) for which CO and \w{} data have since become available \citep{kristensen10b,yildiz10a}. The three test cases are all nearby protostars ($d=180$--450 pc) and cover the evolutionary range from deeply embedded Stage 0 to late Stage I.

The question of the origin of the hot gas observed with \emph{Herschel} is not limited to embedded low-mass young stellar objects (YSOs). \emph{Herschel} observations of rotationally excited CO and/or \w{} include intermediate-mass YSOs \citep{fich10a,johnstone10a}, high-mass YSOs \citep{chavarria10a}, circumstellar disks \citep{meeus10a,sturm10a}, the \object{Orion Bar} photon-dominated region \citep{habart10a}, and even extragalactic sources like the ultraluminous infrared galaxy \object{Mrk~231} \citep{vanderwerf10a}. A quantitative explanation of the origin of the observed emission is still missing in most of these cases. Our model may help in constraining possible excitation mechanisms in environments other than low-mass YSOs. In particular, the treatment of how the gas is excited by UV fields and shocks can be adapted for different types of sources.

This work is part of the two complementary \emph{Herschel} Key Programmes WISH and DIGIT\@. The primary goal of WISH, or ``Water in star-forming regions with \emph{Herschel}'' \citep{vandishoeck11a}, is to use \w{} as a probe of the physical and chemical conditions during star formation and to follow its abundance as a function of evolutionary phase. DIGIT, or ``Dust, ice, and gas in time'' (PI: N.\ Evans), aims to study the evolution of the gas during different stages of star formation, as well as that of the dust grains and their icy mantles. Observations of CO are an integral part of both programmes, because its simple chemistry and easy use in radiative transfer codes make CO an excellent tool to constrain physical conditions and processes. Indeed, the approach in the current paper is to fit the free parameters in the model to the CO data and then to assess how well the model reproduces the \w{} data. More generally, the model is ultimately intended as a tool in using the \emph{Herschel} data as a probe of the energetics and dynamics of star formation, in particular how and where the young star deposits energy back into the parent molecular cloud. Future work in WISH and DIGIT will explore these questions in more detail.

A detailed description of the model appears in Sect.\ \ref{sec:mdesc}. The \emph{Herschel} data and complementary ground-based data are summarised in Sect.\ \ref{sec:obs}. The model results for CO and \w{} are presented in Sect.\ \ref{sec:res} and discussed in Sect. \ref{sec:disc}. Conclusions are drawn in Sect. \ref{sec:conc}.


\section{Model description}
\label{sec:mdesc}
\begin{figure}[t]
\centering
\includegraphics[width=\hsize]{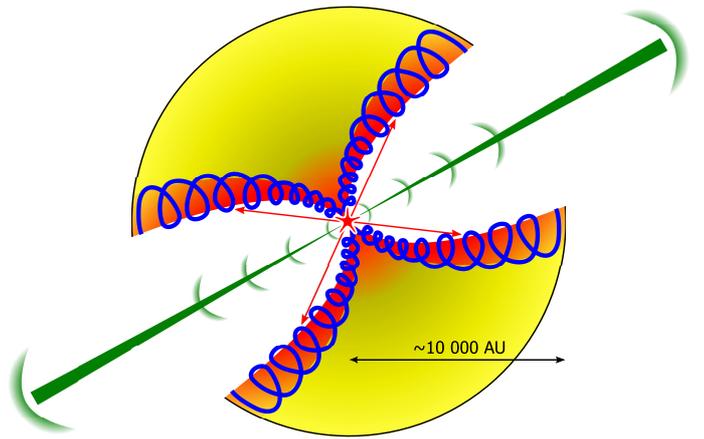}
\caption{Schematic view of the physical components in an embedded Stage 0 or I young stellar object: a passively heated envelope (yellow, with a hot core shaded red), a bipolar jet (green, with bow shocks and internal working surfaces), UV-heated outflow cavity walls (red), and small-scale shocks along the cavity walls (blue). Not visible on this scale is an embedded circumstellar disk with a radius of $\sim$100 AU.}
\label{fig:cartoon}
\end{figure}

The far-infrared (far-IR) and sub-millimetre (sub-mm) molecular line emission from embedded low-mass protostars has traditionally been interpreted with purely spherical envelope models, but such models cannot reproduce the observed intensities of rotationally excited lines like CO 6--5 and higher \citep[e.g.,][]{vankempen09b}. Our model goes a step further by introducing a bipolar outflow cavity with two additional heating mechanisms: UV irradiation of the gas in the cavity walls and small-scale C-type shocks along the walls (\fig{cartoon}). Rotationally excited emission may also arise from the bipolar jet and the associated internal J-type shocks, from gas heated by X-rays, or from a circumstellar disk embedded within the envelope. However, our goal is not to create a single all-inclusive model of an embedded protostar. Rather, the main question in this work is what the \emph{Herschel} observations can teach us about the passively heated envelope, the UV-heated cavity walls and the small-scale shocks.
Investigating the quantitative contributions from other components will be left for future studies.
Furthermore, the current focus is on the integrated line intensities only. With regards to the \emph{Herschel}-PACS data, we are only interested in the emission from the central spaxel of the full $5\times5$ spaxel array. Future updates of the model will explore the spatial extent of the emission and could add dynamical processes to allow for the computation of line profiles.

\subsection{Passively heated envelope}
\label{subsec:env}
Low-$J$ lines from CO and its isotopologues and similar low-excitation lines from other species are commonly modelled with a spherical envelope \citep[e.g.,][]{jorgensen02a,schoier02a}. A spherical envelope is also the starting point of our model, since the warm inner region can contribute to emission in the higher-$J$ lines. The gas in the envelope is well coupled to the dust and is only heated passively or indirectly, i.e., the dust is heated by the protostellar luminosity and the gas attains the same temperature through collisions with the dust.

The density and temperature profiles of each source are constrained from continuum observations. Following the method of \citet{jorgensen02a} and \citet{schoier02a}, the continuum emission for a grid of envelope models is calculated with the 1D continuum radiative transfer program DUSTY \citep{ivezic99a}. The best-fit model is determined by comparing the emission to spectral energy distributions (SEDs) and sub-mm brightness profiles compiled by \citet{froebrich05a} and \citet{difrancesco08a}. The inner radius of the envelope is set at a dust temperature of 250 K (about 30 AU or $0.15''$ at 200 pc; \tb{modpar}). If any material is present on smaller scales, its contribution to the overall line emission would be too diluted in the $\sim$$10''$ \emph{Herschel} beam to have a significant impact on the model results. The envelopes are optically thick at UV and visible wavelengths, so most of the stellar radiation is reprocessed by the dust. For a constant stellar luminosity, the assumed stellar temperature therefore does not have a significant effect on the resulting dust temperatures or the molecular line emission \citep{jorgensen02a}. This was verified by running part of the DUSTY grid for stellar temperatures of both 5\,000 and 10\,000 K.

The remaining three free parameters are fitted in a $\chi^2$ procedure: the power-law index for the density profile ($n \propto r^{-p}$), the optical depth at 100 \micron{} ($\tau_{100}$), and the size of the envelope ($Y \equiv r\q{out}/r\q{in}$). For our line radiative transfer models, the envelopes are terminated at the point where the density reaches \ten{3} \pcc{} or the temperature reaches 10 K \citep{jorgensen02a}. This point is denoted $r\q{env}$. The best-fit values for the three test cases are listed in \tb{modpar} in Sect.\ \ref{sec:obs}, and the associated density and temperature profiles are plotted in \fig{1denv}.

The Doppler broadening parameter $b$ for the passive envelope models is fixed at 0.8 km \ps{}, a typical value for the quiescent gas in low-mass YSOs as determined from optically thin C$^{18}$O observations \citep{jorgensen02a}. The ongoing collapse of the envelope onto the star and disk is simulated with a power-law infall velocity profile $\varv \propto r^{-1/2}$ \citep{shu77a}, scaled to 2.0 km \ps{} at the inner edge of the envelope. Details on how the abundances are obtained and how the molecular line intensities are calculated are provided in Sects.\ \ref{subsubsec:chem} and \ref{subsubsec:rtran}.


\subsection{Photon-heated cavity walls}
\label{subsec:walls}


\subsubsection{Density and temperature}
\label{subsubsec:denstemp}
The second model component is the UV-heated gas in the walls of the outflow cavities \citep[red in \fig{cartoon};][]{spaans95a,vankempen09b}. For an ambient cavity density of a few \ten{3} \pcc{} \citep{neufeld09a}, the total gas column from the central source to the point where the cavity intersects the outer edge of the envelope (a distance of $\sim$\ten{17} cm) is a few \ten{20} \pcs. This corresponds to a UV optical depth of a few tenths, so UV photons produced near the protostar can freely reach the cavity walls. When they do, they produce a photon-dominated zone where the gas is heated to temperatures of a few hundred K\@.

The first step in modelling the emission from the UV-heated gas is to carve out a bipolar cavity from the spherical envelope, similar to the procedure followed by \citet{bruderer10a}. Mid-infrared \emph{Spitzer Space Telescope} images of HH~46 and similar YSOs like \object{L~1527} show ellipsoidal cavities, with a ratio between the semimajor and semiminor axes ($a\q{cav}$ and $b\q{cav}$) of about 4 \citep{velusamy07a,tobin08a}. Large-scale CO 3--2 maps of NGC1333 IRAS2A, tracing the swept-up outflow gas, show roughly the same shape \citep{sandell94a}. DK~Cha is oriented close to pole-on \citep{knee92a}, so the shape and size of its outflow are unknown.

Tests of our model show that the variation in the computed CO and \w{} line fluxes resulting from changes in the shape and size of the cavities is within the overall model uncertainties, consistent with the more detailed tests of \citet{bruderer10a}. Our approach is therefore to fix the ratio between the outflow axes ($a\q{cav}/b\q{cav}$) of all three sources to the value of 4.3 measured for HH~46 \citep{velusamy07a}, and to scale the absolute length of the axes with the envelope radius. In other words, the ratios $a\q{cav}/r\q{env}$ and $b\q{cav}/r\q{env}$ of all three sources are fixed to the values of 2.1 and 0.50 derived for HH~46. The envelope radii and outflow axis lengths are summarised in \tb{modpar}.

\begin{figure}[t]
\centering
\includegraphics[width=\hsize]{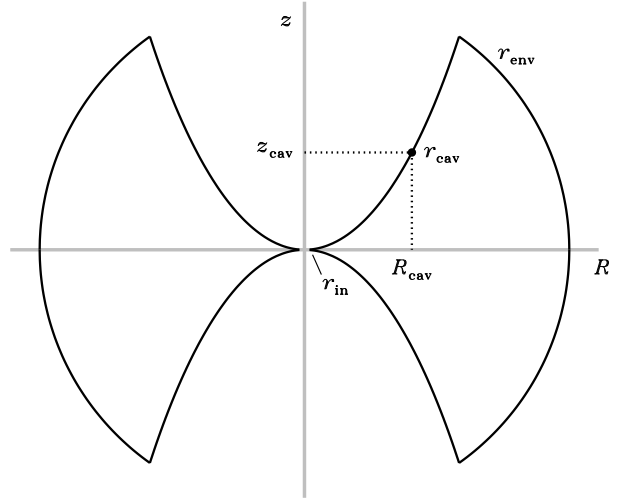}
\caption{Schematic view of the envelope and the bipolar outflow cavity. In each quadrant, a point on the cavity wall is uniquely identified by its distance $r\q{cav}$ to the central star at the origin (\eq{rcav}).}
\label{fig:extinc}
\end{figure}

The two cavities for each source are aligned along the $z$ axis, with the ends meeting at $z=0$ (\fig{extinc}). The ellipsoidal cavity wall in the upper right quadrant of a cylindrical coordinate system is defined as
\begin{equation}
\label{eq:cavwall}
R\q{cav} = b\q{cav}\sqrt{1-\left(\frac{z\q{cav}}{a\q{cav}}-1\right)^2}\,.
\end{equation}
The points along the wall can also be characterised by their distance $r\q{cav}$ to the protostar,
\begin{equation}
\label{eq:rcav}
r\q{cav} = \sqrt{R^2\q{cav}+z^2\q{cav}}\,,
\end{equation}
which is used for some figures in Sect.\ \ref{sec:res}.

The only UV source included in our model is the protostar. UV photons can also be produced at the bow shock \citep{bohm93a,raymond97a} or by small-scale shocks within the cavity or along the walls \citep{curiel95a,saucedo03a,walter03a}, but these contributions are highly uncertain. The protostellar UV luminosity itself is also highly uncertain, and is therefore treated as a free parameter.
Classical T~Tauri stars have typical $L\q{UV}$ of \ten{-3.5}--\ten{-1.5} times the total stellar luminosity \citep{ingleby11a}. Scaling up to the higher accretion rates found in embedded YSOs, anything from 0.01 to 1 $L_\odot$ would be a reasonable value for our model. The best-fit UV luminosities derived in Sect.\ \ref{subsec:chisq} range from $<$0.03 to 0.08 $L_\odot$.
An $L\q{UV}$ of 0.1 $L_\odot$ would give an unattenuated UV flux at 100 AU of $G_0\approx10^4$ relative to the mean interstellar radiation field of \citet{habing68a}.

Due to the UV field impinging on the cavity walls, the gas is heated to higher temperatures than it attains in the passively heated, purely spherical envelope. The temperature can be calculated by solving the heating and cooling balance as function of the physical conditions. Many such models are available in the literature, but they show substantially different results, sometimes even at the qualitative level \citep{rollig07a}. The consequences of these uncertainties for our model results are discussed extensively in Sect.\ \ref{subsec:tgas}. For the initial results in Sect.\ \ref{sec:res}, the PDR surface temperature grid of \citet{kaufman99a} is adopted as our standard prescription. This grid was calculated for densities up to \ten{7} \pcc; for higher densities, it is extrapolated as $1/n(\mhm)$ based on the expected dominant heating and cooling mechanisms (Sect.\ \ref{subsec:tgas}).

The \citet{kaufman99a} grid as such can only be used at the cavity wall. At any depth into the envelope, the temperature needs to be corrected for attenuation of the UV field by dust. \citet{rollig07a} plotted the temperature as function of the visual extinction, $\av$, for several densities and unattenuated UV fluxes. In each case, the temperature decreases roughly as $\exp(-0.6\av)$. In order to use this depth dependence in our model, the amount of extinction towards any point ($R,z$) is calculated with the continuum radiative transfer code RADMC \citep{dullemond04a}. The input stellar spectrum is a 10\,000 K blackbody scaled to a given UV luminosity between 6 and 13.6 eV\@. This input spectrum is a reasonable approximation of the real spectrum emitted by a low-mass protostar with a UV excess due to accretion \citep[e.g.,][]{vanzadelhoff03a}. The local UV flux from RADMC, $F\q{RADMC}$ (also integrated from 6 to 13.6 eV), can be considered the product of the unattenuated flux and an extinction term,
\begin{equation}
\label{eq:fradmc}
F\q{RADMC}(R,z) = F\q{unatt} \exp(-\tauuv)\,,
\end{equation}
where $F\q{unatt}$ is the UV flux through a unit surface arising from a source at a distance $\sqrt{R^2+z^2}$. The UV optical depth, $\tauuv$, is effectively an average over the many possible paths a photon can follow from the source to the point ($R,z$). It is converted into a visual extinction via $\av=\tauuv/3.02$ \citep{bohlin78a}. Along with the UV field, RADMC also calculates the dust temperature along the cavity wall and throughout the envelope. This is used instead of the dust temperature from the 1D DUSTY models.

One caveat with RADMC is that it treats scattering of the radiation by dust as an isotropic process, while in reality grains are partially forward scattering \citep{li01a}. Tests with the radiative transfer code of \citet{vanzadelhoff03a}, which does allow for anisotropic scattering, show flux differences of at most 20\% between an isotropic scattering function and the scattering function of \citeauthor{li01a}. This is within our overall model uncertainty.

The present goal is to reproduce the observed integrated line fluxes, not the detailed line profiles. Hence, the Doppler broadening and infall velocities for the UV-heated component are kept the same as in the passively heated component.


\subsubsection{Abundances}
\label{subsubsec:chem}
Several thousand grid points are required to properly sample the envelope and the cavity walls in the 3D radiative transfer calculations (Sect.\ \ref{subsubsec:rtran}). It would be too time consuming to calculate the abundances of CO and \w{} for each individual point with a traditional chemical code. Instead, the abundances are obtained according to the method of \citet{bruderer09a}. A grid of chemical models is pre-computed for the appropriate range of densities, temperatures, unattenuated UV fluxes and extinctions. Abundances for each individual point in the envelope or the cavity wall are interpolated from this grid when constructing the source models for the radiative transfer code.

The basis of our chemical reaction network is the UMIST06 database \citep{woodall07a} as modified by \citet{bruderer09a}, except that X-ray chemistry is not included. The network allows all neutrals to freeze out onto the dust to account for the depletion of many gas-phase species in the colder parts of the envelope. Molecules can return to the gas phase through thermal desorption and photodesorption. Binding energies and photodesorption yields for CO and \w{} are taken from \citet{fraser01a}, \citet{bisschop06a} and \citet{oberg07a,oberg09a}. Photodesorption occurs even in regions of high extinction because of a cosmic-ray--induced UV flux of about \ten{4} photons \pcs{} \ps{}, equivalent to \ten{-4} times the mean interstellar flux \citep{shen04a}. In the gas phase, UV photons can dissociate or ionise many species; the relevant rates are calculated according to \citet{vandishoeck06a} for a 10\,000 K blackbody stellar spectrum, appropriate for a low-mass protostar with a UV excess. Self-shielding of \mh{} and CO, which causes their dissociation rates to decrease more rapidly than those of other species, is included according to \citet{draine96a} and \citet{visser09b}. The cosmic-ray ionisation rate of \mh{} is set to \scit{5}{-17} \ps{} \citep{dalgarno06a}.

The grid of pre-computed abundances, from which abundances for the three sources are interpolated, is obtained by advancing the chemical reactions for a period of $10^5$ yr. As starting conditions, all carbon is locked up in CO, the remaining oxygen in \w, all nitrogen in \mn, and all remaining hydrogen in \mh. Reflecting the most likely formation pathways, CO and \mn{} start in the gas phase, while \w{} starts frozen out on the dust. \mh{} is predominantly formed on the grains, but because it rapidly evaporates even at 10 K, it starts in the gas phase. Elemental abundances are adopted from \citet{aikawa08a} and \citet{bruderer09a}; the initial
composition is summarised in \tb{initabun}.

\begin{table}
\caption{Initial abundances of molecules in our chemical grid.}
\label{tb:initabun}
\centering
\begin{tabular}{lc}
\hline\hline
\cellcent{Species}       & Abundance \\
\hline
\mh                      & 1.00(0)\phantom{$-$} \\
CO                       & 1.57($-$4) \\
H$_2$O$\q{ice}$       & 2.03($-$4) \\
\mn                      & 4.94($-$5) \\
\hline
\end{tabular}
\tablefoot{The notation $a(b)$ means \scit{a}{b}. Abundances are given relative to \mh. Initial abundances of elements present in the UMIST06 database but not listed in this table are as in \citet{aikawa08a} and \citet{bruderer09a}.}
\end{table}


\subsubsection{Radiative transfer}
\label{subsubsec:rtran}
Level populations and line intensities for the passively heated envelope and the UV-heated outflow cavity walls are calculated with the new full-3D radiative transfer code LIME \citep{brinch10a}. LIME uses an irregular grid with points sampled randomly but weighted according to a user-defined criterion such as density or temperature.
This naturally results in a finely spaced grid in the regions most important for the astrophysical problem at hand.

The source geometry of our models introduces a substantial difficulty: the high-excitation CO and \w{} lines originate in a thin layer of hot gas along the cavity wall, so an accurate flux calculation requires a high grid point density inside and immediately outside this small region. Furthermore, the grid has to be able to resolve emission on scales ranging from $\sim$10 to $\sim$\ten{4} AU\@. This cannot be accomplished with the default density or temperature weighting functions. Extensive convergence tests were carried out to find an alternative grid sampling scheme. The optimised grid consists of 30\,000 points, picked randomly in $\log{r}$ between $r\q{in}$ and $r\q{env}$. For every point with radial coordinate $r$, a point ($R\q{cav},z\q{cav}$) is located on the cavity wall such that $\sqrt{R\q{cav}^2+z\q{cav}^2}=r$. Let $\gamma\q{cav}$ be the angle between the point ($R\q{cav},z\q{cav}$) and the midplane, i.e., $\tan{\gamma\q{cav}}=z\q{cav}/R\q{cav}$. The angular coordinate $\gamma$ of each point is then determined as follows:
\begin{enumerate}
\item one third of the points are placed exactly at the cavity wall ($\gamma=\gamma\q{cav}$);
\item one third of the points are picked with a random angle $\gamma$ such that $0.9<\gamma/\gamma\q{cav}\leq1$;
\item one third of the points are picked with a random angle $\gamma$ such that $0<\gamma/\gamma\q{cav}\leq1$ and then weighted by $\sqrt{\gamma/\gamma\q{cav}}$, which favors points near the midplane over points near the cavity wall.
\end{enumerate}
The grid thus constructed is smoothed by five iterations of Lloyd's algorithm \citep{lloyd82a}, which moves each point somewhat towards the centre of mass of its Voronoi cell. The purpose of this smoothing step is two-fold: (1) to turn some ``needle-shaped'' cells created by the random setup into more regular shapes \citep{brinch10a}; and (2) to push some grid points from the first sampling group into the cavity wall, so that the edge between the envelope and the cavity is well sampled on either side. The first two groups of points together cover the region of hot gas responsible for the more excited lines. The third group of points covers the bulk of the envelope, where the low-$J$ emission originates.

With the grid established, LIME employs the standard two-step method for problems where local thermodynamic equilibrium (LTE) does not apply. The first step is an iterative procedure to calculate the rotational level populations based on the balance between radiative excitation, collisional excitation and collisional de-excitation. The second step is the ray tracing, i.e., calculating the emergent spectrum at a given distance and inclination. The collisional rate coefficients required for the first step are collected from the Leiden Atomic and Molecular Database\footnote{http://www.strw.leidenuniv.nl/$\sim$moldata} \citep[LAMDA;][]{schoier05a}, where the most recent data files for CO and \w{} are from \citet{yang10a} and \citet{faure07a}, respectively. The ortho-to-para ratio for \mh{} is taken to be thermalised, with a maximum value of 3. The ortho-to-para ratio of \w{} is fixed at 3. Dust is included in the model with a standard gas-to-dust ratio of 100 and the OH5 opacities from \citet{ossenkopf94a}, appropriate for dust grains with a thin ice mantle.

Images are created at 0\farcs1 resolution (18--45 AU at the distances of the three sources; \tb{modpar}) to get a sufficiently fine sampling of the hot inner regions. In order to compare the model results to the observations, each image is convolved with a Gaussian beam profile. Beam sizes are calculated as the diffraction-limited beam for the wavelength and the telescope (\emph{Herschel}, APEX or JCMT) at which the simulated line was observed. In the case of lines observed with \emph{Herschel}-PACS, fluxes are extracted from the central $9\farcs4\times9\farcs4$ to mimic the instrument's central spaxel. For a given density, temperature and abundance structure, the estimated uncertainty on the CO line fluxes ranges from 30\% for the 2--1 line to a factor of 2 for the high-$J$ end of the ladder. For \w, it is a factor of 2 for all lines. The uncertainty is mainly due to the complex geometry and the small size of the emitting region for the high-$J$ lines. The source models themselves lead to uncertainties of similar magnitude in the line fluxes, mostly because of the problems with the gas temperature discussed in Sect.\ \ref{subsec:tgas}.

Integrated intensities from APEX, the JCMT and \emph{Herschel}-HIFI ($\tmbdv$ in \kkms) and integrated fluxes from \emph{Herschel}-PACS ($\fnudnu$ in W m$^{-2}$) can be mutually converted through
\begin{equation}
\label{eq:hifi2pacs}
\fnudnu = \frac{2k\Omega}{\lambda^3}\tmbdv\,,
\end{equation}
with $\Omega(\lambda)$ the solid angle subtended by the beam at the given wavelength. Denoting the full width at half maximum (FWHM) of the beam as $\Theta(\lambda)$, the solid angle for APEX, the JCMT and HIFI is $(\pi/4\ln2)\Theta^2$. The formula is more complicated for PACS, because the size of the spaxels has to be taken into account. For just the central spaxel, the solid angle is
\begin{equation}
\label{eq:omegapacs}
\Omega\q{PACS}(\lambda) = \left[\frac{9\farcs4}{{\rm erf}\left(9\farcs4\sqrt{\ln{2}}/\Theta\right)}\right]^2 \,,
\end{equation}
with erf the error function and $\Theta(\lambda)$ the FWHM of the diffraction-limited \emph{Herschel} beam.


\subsection{Small-scale shocks}
\label{subsec:cshock}
The third model component is a series of C-type shocks moving along the cavity walls. These shocks may be powered by the wind launched from close to the protostar or the surface of the inner disk, or by the expansion of the jet \citep[e.g.,][]{lada85a, shang06a, arce06a, santiago09a}. Shocks are created when the wind hits the outflow cavity walls, heating the gas to temperatures of more than 1000 K.

The model procedure is the same as that of \citet{kristensen08a}, except for a change in the geometry. In the original procedure, 1D shock model results were pasted along a parabola to simulate emission from a bow shock observed in \mh{} in the \object{Orion Molecular Cloud}. Here, in order to simulate C-type shocks along the cavity walls, the 1D shocks are pasted along the ellipsoids from Sect.\ \ref{subsubsec:denstemp}. The shock velocity is assumed to be constant along the entire length of the wall and the pre-shock density is that of the envelope at the given distance from the protostar. The magnetic field is set to $\beta[2n(\mhm)/{\rm cm}^{-3}]^{1/2}$, with $\beta$ fixed at 1 $\mu$Gauss. Abundances as function of depth into the shock are calculated from a limited chemical network centred on \w{} \citep{wagner87a}, which does not include photodissociation or other UV-driven reactions.

\begin{figure}[t]
\centering
\includegraphics[width=\hsize]{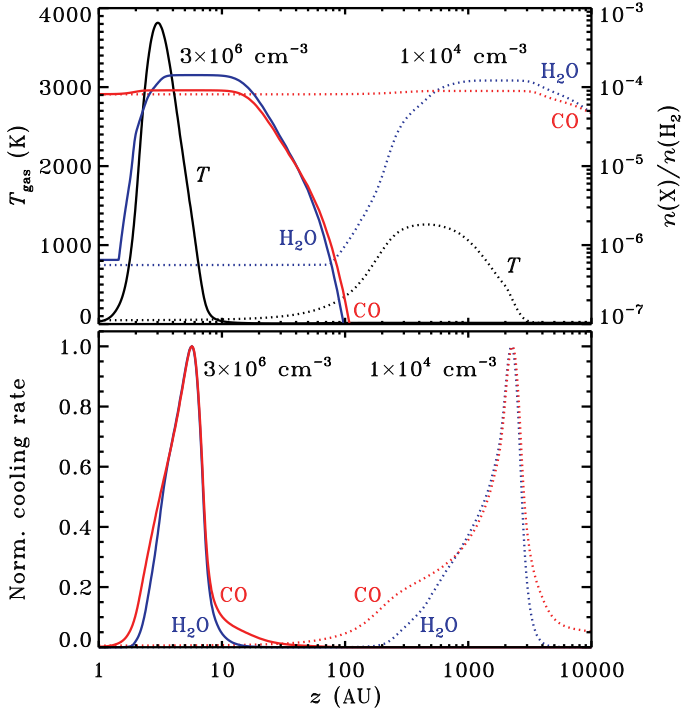}
\caption{\emph{Top:} Temperature of the neutral gas (black) and abundances of CO (red) and \w{} (blue) as function of depth into a shock for a shock velocity of 20 km \ps{} and pre-shock densities of \scit{3}{6} (solid) and \scit{1}{4} \pcc{} (dotted). \emph{Bottom:} Normalised cooling rates of CO and \w{} for the same two sets of conditions. All data are taken from the models of \citet{flower03a}.}
\label{fig:cool1}
\end{figure}

The shock model results of \citet{kaufman96a} are used in combination with the model results of \citet{flower03a} and Kristensen et al.\ (in prep.) to compute the emergent line intensities. The first step is to estimate the extent of the line-emitting region by measuring the FWHM of the cooling-rate profile of the relevant molecule through the shock. The bottom panel of \fig{cool1} shows the cooling profiles from \citeauthor{flower03a} for CO and \w{}, for a shock velocity of 20 km \ps{} and pre-shock densities of \ten{4} and \ten{6.5} \pcc. The corresponding temperature and abundance profiles are plotted in the top panel. CO and \w{} cool the same part of the shock and their cooling lengths are nearly the same. Furthermore, the cooling lengths are nearly inversely proportional to the pre-shock density, i.e., the solid and dotted profiles look similar but are shifted on the logarithmic scale. The cooling lengths provide an estimate of the extent over which the shock is emitting, and as such are an effective measure of the beam-filling factor.

\begin{figure}[t]
\centering
\includegraphics[width=\hsize]{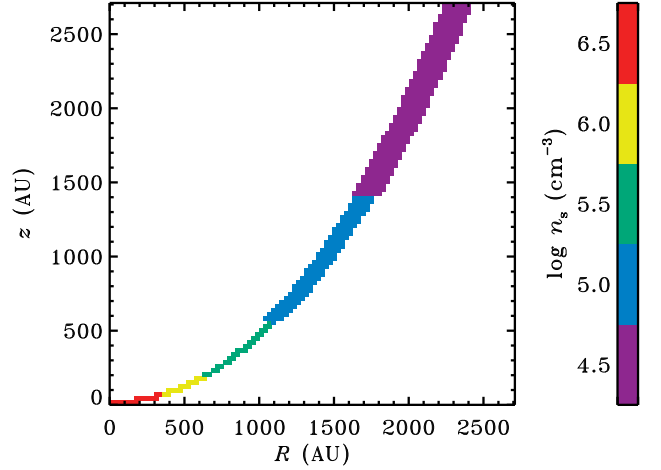}
\caption{Segmentation of the outflow cavity wall of HH~46 into regions of different pre-shock densities. The cooling length is larger for lower densities, so the shocks are wider at larger radii.}
\label{fig:segment}
\end{figure}

The second step is to divide the cavity wall into discrete segments according to the envelope density.
The upper limit of the \citet{kaufman96a} grid is \ten{6.5} \pcc; the pre-shock density for the denser parts of the envelope is set to that value. \figg{segment} shows the adopted segmentation for HH~46; the width of the coloured area is a rough representation of the size of the emitting region. The CO and \w{} line intensities are estimated from the results tabulated by \citeauthor{kaufman96a}. The emission is assumed to fill each of the segments, thus creating an emission map for each line. Just like the raw images produced by LIME (Sect.\ \ref{subsubsec:rtran}), the maps are convolved with a Gaussian beam profile of the appropriate diameter. For lines observed with \emph{Herschel}-PACS, the flux is extracted from the central $9\farcs4\times9\farcs4$. The fluxes are not corrected for extinction by dust in the envelope.

\begin{table*}
\caption{Observational characteristics, model parameters and derived properties for the three sources.}
\label{tb:modpar}
\centering
\begin{tabular}{lcccccccccccccccc}
\hline\hline
\cellcent{Source} & $d$  & $L\q{bol}$ & $Y$    & $p$ & $\tau_{100}$ & $r\q{in}$ & $r\q{env}$ & $n(r\q{in})$ & $n(r\q{env})$ & $M\q{env}$ & $a\q{cav}$ & $b\q{cav}$ & $i$        & $L\q{UV}$ & $\varv\q{s}$ \\
                  & (pc) & ($L_\odot$)   &        &     &              & (AU)         & (AU)          & (\pcc)          & (\pcc)           & ($M_\odot$)   & (AU)          & (AU)          & ($^\circ$) & ($L_\odot$)  & (km \ps) \\
\hline
IRAS2A            & 235  & 37            & \pz500 & 1.5 & 0.4          & 28.0         &   14\,000     & 1.5(8)          & 1.3(4)           & 2.0\pz        & 29\,000       &    7\,000     &   90       & $<$0.03\phantom{$<$}      & 15 \\
HH~46             & 450  & 26            &   3000 & 2.0 & 1.9          & 34.6         &   16\,100     & 1.1(9)          & 5.0(3)           & 1.7\pz        & 34\,000       &    8\,000     &   53       & 0.08      & 15 \\
DK~Cha            & 178  & 36            &   3000 & 2.1 & 0.1          & 25.6         & \pz5\,700     & 8.5(7)          & 1.0(3)           & 0.02          & 12\,000       &    2\,800     & \pz0       & 0.05        & $>$20\phantom{$>$} \\
\hline
\end{tabular}
\tablefoot{References: Kristensen et al.\ (subm.) for $L\q{bol}$; \citet{sandell94a} and \citet{hirota08a} for NGC1333 IRAS2A; \citet{graham89a}, \citet{velusamy07a} and \citet{nishikawa08a} for HH~46; \citet{knee92a} and \citet{whittet97a} for DK~Cha.}
\end{table*}

Our method is specifically designed to simulate optically thin emission from shocks. Low-$J$ CO lines typically observed in and associated with outflows, such as the 2--1 and 3--2, originate in cold ($\sim$100 K) swept-up gas present on larger spatial scales than the shocks along the cavity wall \citep[e.g.,][]{arce07a}. These low-$J$ lines are often optically thick and only probe the surface layer of the outflow, where the velocity gradient is largest.
Since our focus is on the hot gas observed with \emph{Herschel}, no attempt is made to quantify the emission from the colder swept-up gas.

An important caveat in the shock models is that they were calculated without a UV field. Several groups are working on irradiated shock models featuring a self-consistent treatment of the UV field in both the pre- and the post-shock gas (e.g., \citealt{lesaffre11a}; M.~J.\ Kaufman, priv.\ comm.), but no such models are currently available. Qualitatively, a UV-irradiated shock is expected to be smaller and hotter, and to have a lower \w{} abundance. The latter is a direct effect of the photodissociation of \w\@. The smaller spatial extent and the higher peak temperature are due to the higher degree of ionization in the irradiated pre-shock gas: the ion-neutral coupling length decreases, so the shock front is compressed and the same amount of energy is imparted over a shorter distance. The shock results are discussed in Sect.\ \ref{sec:res} with these qualitative arguments in mind.


\section{Source sample and observations}
\label{sec:obs}
The three test cases for the model are the low-mass YSOs NGC1333 IRAS2A, HH~46 IRS and DK~Cha (IRAS 12496$-$7650). Each source has been studied extensively in the past, including recent observations with PACS and HIFI on \emph{Herschel} \citep{vankempen10b,vankempen10a,kristensen10b,yildiz10a}. The availability of \emph{Herschel} data and a large body of complementary data is one reason to choose these three sources. Another is that they offer various challenges to the model by spanning a range of evolutionary stages and orientations: IRAS2A is a deeply embedded Stage 0 source seen nearly edge-on \citep{sandell94a}, HH~46 is a Stage I source seen at a 53$^\circ$ inclination \citep{nishikawa08a}, and DK~Cha is in transition from Stage I to II and is seen nearly pole-on \citep{knee92a}.
The basic observational characteristics and model parameters for each source are summarised in \tb{modpar}.

Far-IR spectra of the three sources were obtained with PACS and HIFI on \emph{Herschel} \citep{pilbratt10a,poglitsch10a,degraauw10a}.
The HIFI data used in this work were published by \citet{kristensen10b} and \citet{yildiz10a}. The PACS spectra of HH~46 and DK~Cha were published by \citet{vankempen10b,vankempen10a}; the PACS spectrum of IRAS2A is previously unpublished. All PACS data were rereduced with the standard pipeline in HIPE v6.1 \citep{ott10a}. A spectral flatfield was applied to the data to improve the signal-to-noise ratio.

PACS consists of a 5$\times$5 integral field unit with $9\farcs4\times9\farcs4$ spaxels. For HH~46, the fluxes of the central 9\farcs4 region are measured directly from the central spaxel and are accurate to 10--20\%. The observations of DK~Cha and IRAS2A were mispointed by a few arcsec. For both sources, the line emission is concentrated near the location of the continuum emission. In order to measure the line fluxes, the spectra in the four spaxels with the brightest continuum emission were first summed. The ratio of the continuum emission in these four spaxels to the total continuum flux in the $5\times5$ array was then used to obtain a wavelength-dependent curve to convert the four-spaxel extraction into a flux in the entire field of view. The fluxes were multiplied by the fraction of the flux that falls in the central spaxel (ranging from 70\% below 100 \micron{} to 40\% at 200 \micron) to estimate the total flux that would have been in the central spaxel of a well-pointed observation. These corrections assume a point source and may therefore overestimate the flux in the central spaxel. Excluding this uncertainty, the fluxes for IRAS2A and DK~Cha are accurate to about 30\%. Fluxes and upper limits for all three sources are tabulated in \tb{lineobs}.

\begin{table}
\caption{Measured line fluxes (\ten{-18} W m$^{-2}$) from the central region around each source.}
\label{tb:lineobs}
\centering
\begin{tabular}{cccccc}
\hline\hline
Transition & $\eup/k$ & $\lambda$  & IRAS2A & HH~46 & DK~Cha\\
           & (K)      & (\micron)  &        &       & \\
\hline
CO & & & & & \\
14--13 & 581 & 186.00 & 73 & 52 & 207 \\
15--14 & 663 & 173.63 & 112 &  & 221 \\
16--15 & 752 & 162.81 & 67 & 44 & 219 \\
17--16 & 846 & 153.27 & 64 &  & 238 \\
18--17 & 945 & 144.78 & 80 & 58 & 245 \\
19--18 & 1050 & 137.20 & 95 &  &  \\
20--19 & 1160 & 130.37 & $<$51 &  &  \\
21--20 & 1276 & 124.19 & $<$94 &  &  \\
22--21 & 1397 & 118.58 & $<$53 & 35 & 186 \\
\textit{23--22} & \textit{1524} & \textit{113.46} & \textit{177} & \textit{74} & \textit{327} \\
24--23 & 1657 & 108.76 & 62 & 26 & 200 \\
27--26 & 2086 & 96.77 & 187 &  &  \\
28--27 & 2240 & 93.35 & 140 &  &  \\
29--28 & 2400 & 90.16 & $<$35 & 20 & 254 \\
30--29 & 2565 & 87.19 & $<$50 & 18 & 147 \\
\textit{31--30}	& \textit{2735}	& \textit{84.41} & \textit{297} & \textit{74} & \textit{263} \\
32--31 & 2911 & 81.81 & $<$33 & $<$33 & 163 \\
33--32 & 3092 & 79.36 & $<$36 & $<$23 & 92 \\
34--33 & 3279 & 77.06 & $<$122 &  & 234 \\
35--34 & 3471 & 74.89 & $<$29 &  & 209 \\
36--35 & 3669 & 72.84 & $<$131 & $<$24 & 129 \\
37--36 & 3872 & 70.91 & $<$33 &  & 69 \\
38--37 & 4080 & 69.07 & $<$25 &  & 118 \\
\hline
\w & & & & & \\
$2_{21}$--$2_{12}$ & 194 & 180.49 & 77 & 6 & $<$23 \\
$2_{12}$--$1_{01}$ & 114 & 179.53 & 105 & 7 & 33 \\
$3_{03}$--$2_{12}$ & 197 & 174.62 & 101 & 10 & 89 \\
$3_{22}$--$3_{13}$ & 297 & 156.19 & 109 &  &  \\
$3_{13}$--$2_{02}$ & 205 & 138.53 & 105 & 8 & 69 \\
$4_{04}$--$3_{13}$ & 320 & 125.35 & 58 & 4 & 38 \\
$5_{33}$--$5_{24}$ & 725 & 113.95 & $<$33 & $<$6 & $<$31 \\
$\mathit{4_{14}}$--$\mathit{3_{03}}$ & \textit{3234} & \textit{113.54} & \textit{177} & \textit{74} & \textit{327} \\
$7_{43}$--$7_{34}$ & 1340 & 112.51 & 34 &  &  \\
$2_{21}$--$1_{10}$ & 194 & 108.07 & 237 & 13 & $<$74 \\
$7_{44}$--$7_{35}$ & 1335 & 90.05 & $<$33 & $<$5 & $<$63 \\
$3_{22}$--$2_{11}$ & 297 & 89.99 & 97 & $<$5 & $<$63 \\
$7_{16}$--$7_{07}$ & 1013 & 84.77 & $<$29 & $<$7 & $<$45 \\
$6_{06}$--$5_{15}$ & 643 & 83.28 & 182 &  & $<$51 \\
$8_{35}$--$7_{44}$ & 1511 & 81.69 & $<$28 & $<$6 & $<$126 \\
$9_{27}$--$9_{18}$ & 1729 & 81.40 & $<$29 & $<$5 & $<$125 \\
$4_{23}$--$3_{12}$ & 432 & 78.74 & 38 & $<$7 & $<$86 \\
$3_{21}$--$2_{12}$ & 305 & 75.38 & 126 &  & $<$87 \\
$9_{37}$--$9_{28}$ & 1750 & 73.61 & 158 &  & $<$126 \\
$8_{18}$--$7_{07}$ & 1071 & 63.32 & $<$39 & $<$11 & $<$177 \\
\hline
\end{tabular}
\tablefoot{Typical uncertainties are 30\% for IRAS2A and DK~Cha and 10--20\% for HH~46. Upper limits are given at the 3$\sigma$ level. Lines in italics are blended (CO 31--30 with an OH line) and their fluxes are treated as upper limits. Since the observations are compared with model images that are convolved with the \emph{Herschel} point-spread function, the fluxes in this table are non-PSF-corrected.}
\end{table}

The \emph{Herschel} data are combined with observations of lower-$J$ CO and CO isotopologue lines (up to 7--6) obtained with the \emph{James Clerk Maxwell Telescope} (JCMT) and the \emph{Atacama Pathfinder Experiment} \citep[APEX;][]{jorgensen02a,vankempen06a,vankempen09b,yildiz10a}. The different beam sizes for the JCMT, APEX and \emph{Herschel} are taken into account by convolving the line fluxes from the model with a wavelength-dependent beam size appropriate for the instrument with which each line was observed.


\section{Results}
\label{sec:res}


\subsection{Physical and chemical structure}
\label{subsec:physchem}
\begin{figure}[t]
\centering
\includegraphics[width=\hsize]{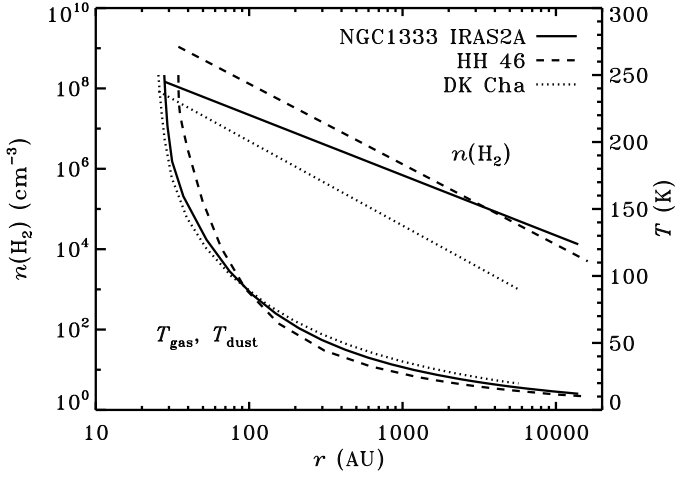}
\caption{Density and temperature profiles for the passively heated envelopes of the three sources.}
\label{fig:1denv}
\end{figure}

The best-fit DUSTY envelope model of HH~46 has a radius of 16\,100 AU (cut off at 10 K) and a mass of 1.7 $M_\odot$ (\tb{modpar}).
The envelope of NGC1333 IRAS2A is a little smaller (14\,100 AU), but because of the shallower density profile its mass comes out 20\% higher at 2.0 $M_\odot$. DK~Cha has the smallest and least massive envelope: 5700 AU and 0.02 $M_\odot$. The density and temperature profiles of each envelope are plotted in \fig{1denv}. 

\begin{figure}[t]
\centering
\includegraphics[width=\hsize]{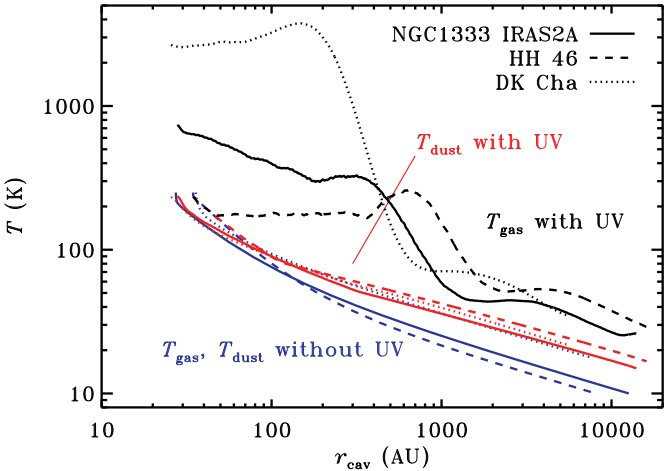}
\caption{Gas temperature (black) and dust temperature (red) along the cavity wall of the three sources due to UV heating. Shocks are not included. The coordinate on the horizontal axis is the distance to the protostar (\eq{rcav}). The blue curves show the temperature profiles due to passive heating only (gas and dust coupled).}
\label{fig:cavtemp}
\end{figure}

\begin{figure*}[t]
\centering
\includegraphics[width=\hsize]{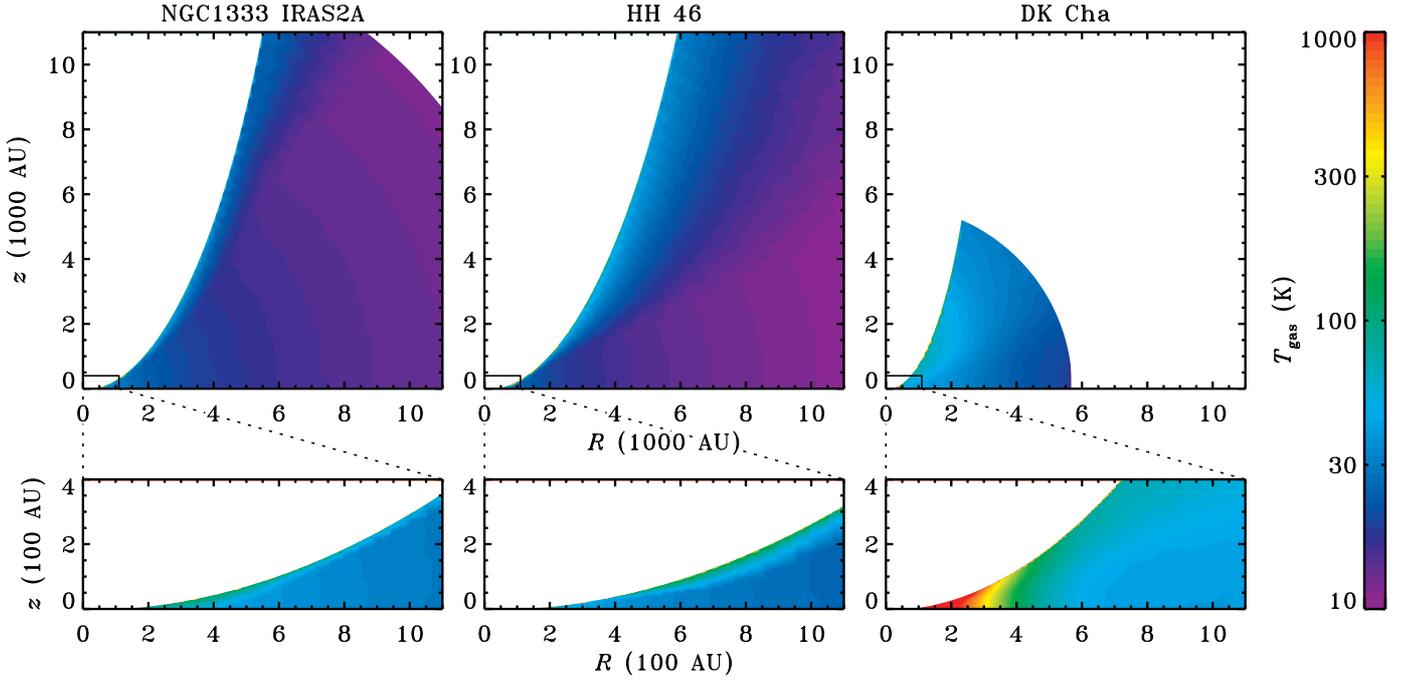}
\caption{Gas temperature profiles due to UV heating of the outflow cavity walls. Shocks are not included. The bottom row zooms in on the regions indicated in the top row.}
\label{fig:tgas}
\end{figure*}

When the cavities are carved out of the spherical envelopes, the gas and dust are heated up by the stellar UV flux. The best-fit UV luminosities for the three central sources are determined in Sect.\ \ref{subsec:chisq} by fitting a grid of luminosities to the CO observations. The resulting temperature profiles along the cavity walls are shown in \fig{cavtemp} (black and red curves). For comparison, the plot also shows the temperature profiles from DUSTY (blue), which is what the gas and dust would reach in the absence of UV heating. The effect of UV heating is strongest for DK~Cha because of the lower envelope densities. The gas temperature peaks at 3800 K at 150 AU from the star. IRAS2A reaches its maximum of 850 K right at the inner edge, followed by a small secondary maximum of 330 K at 320 AU\@. For HH~46, the density in the inner envelope is high enough that the gas initially remains coupled to the dust and has the same temperature. At larger distances, the decreasing density allows the gas to become hotter than the dust, reaching a maximum of 260 K at 670 AU\@.

The full 2D gas temperature profiles in \fig{tgas} show that the effects of UV heating are not limited to just the cavity wall, but penetrate some depth into the envelope according to the $\exp(-0.6\av)$ depth dependence from Sect.\ \ref{subsubsec:denstemp}. The thickness of the UV-heated layer ranges from about 100 AU at $r=1000$ AU to about 2000 AU at the outer edge of the envelope. Hence, the UV-heated layer is about a factor of 10 thicker than the shock-heated layer (\fig{cool1}). Accounting for the predicted effects of the UV field on the shock structure, the thickness ratio between the two layers would be even larger (Sect.\ \ref{subsec:cshock}).

\begin{figure}[t]
\centering
\includegraphics[width=\hsize]{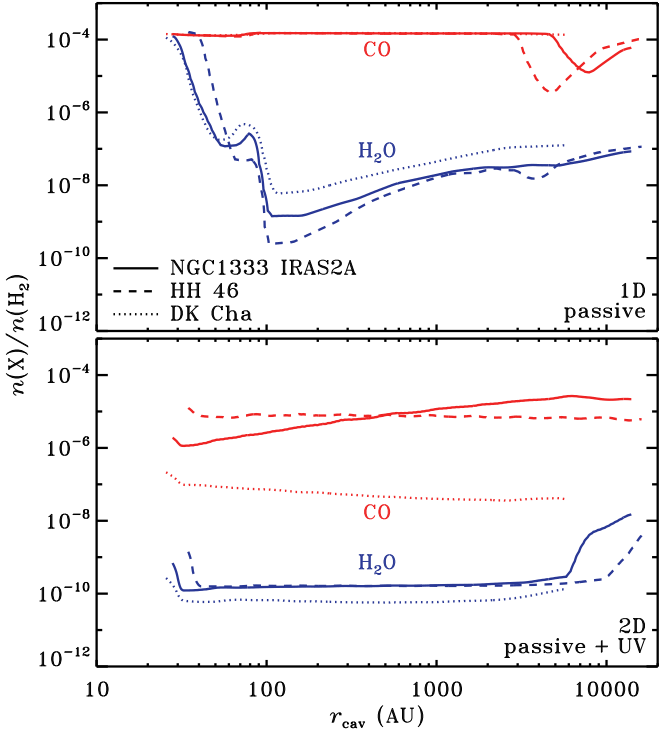}
\caption{\emph{Top:} radial abundance profile of CO (red) and \w{} (blue) in the spherical envelope models with passive heating only. \emph{Bottom:} abundance profiles along the cavity wall (\eq{rcav}) in the 2D models, with all effects of the UV field included: photodissociation, photodesorption, and altered chemistry due to heating of the gas and dust. Shocks are not included in either panel.}
\label{fig:cavabun}
\end{figure}

\begin{figure}[t]
\centering
\includegraphics[width=\hsize]{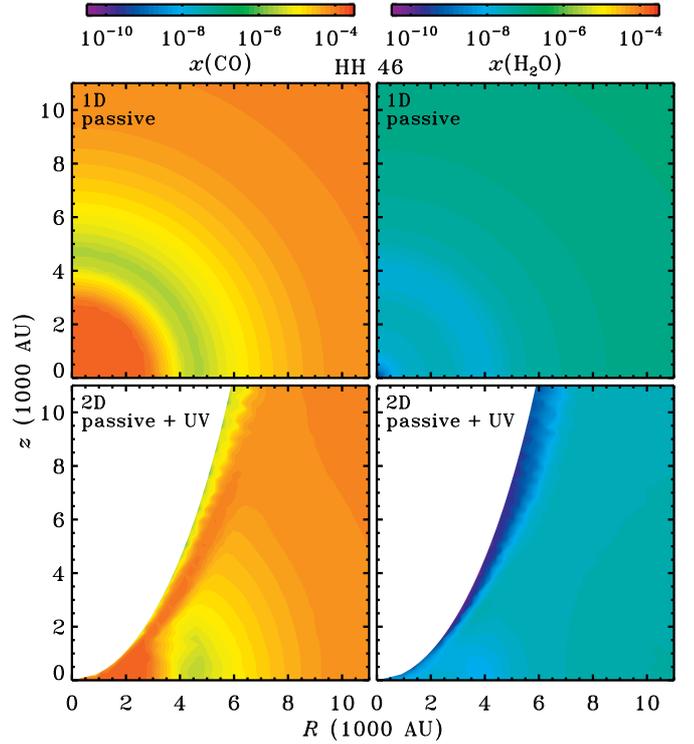}
\caption{Abundance of CO (left) and \w{} (right) in two versions of the HH~46 model: the spherical envelope with passive heating only (top) and the envelope with UV-heated outflow cavity walls (bottom).}
\label{fig:simpleabun}
\end{figure}

In the purely spherical models for IRAS2A and HH~46, the CO abundance follows a drop profile (top panel of \fig{cavabun} and top left panel of \fig{simpleabun}): it is high ($\sim$\ten{-4}) in the warm inner parts of the envelope, low ($\sim$\ten{-6}--\ten{-5}) at larger radii due to freeze-out onto cold dust, and high again ($\sim$\ten{-4}) towards the outer edge because of the low density and corresponding long freeze-out timescale \citep{schoier04a,jorgensen05c,yildiz10a}. DK~Cha has a somewhat warmer envelope (\fig{1denv}), preventing CO from freezing out and keeping its abundance at \ten{-4} at all radii.

There is no freeze-out either along the cavity wall in any of the 2D models (bottom panel of \fig{cavabun} and bottom left panel of \fig{simpleabun}), because the dust temperature now exceeds the CO evaporation temperature everywhere (\fig{cavtemp}). The 10\,000 K blackbody adopted for the stellar spectra is energetic enough to dissociate some CO in the cavity wall, reducing its abundance to \ten{-6}--\ten{-5} in IRAS2A and HH~46, and $\sim$\ten{-7} in DK~Cha (\fig{cavabun}, bottom). Moving from the cavity wall towards the midplane, the increasing amount of extinction shields CO from the dissociating photons, and the abundance goes back up to $\sim$\ten{-4} (\fig{simpleabun}; see \fig{abunco} in the online appendix for the other two sources). The dust near the midplanes of HH~46 and IRAS2A remains cold enough for CO to stay frozen, as evidenced by the green region in the bottom left panel of \fig{simpleabun}.
As a result, horizontal cuts from the cavity wall into the envelope essentially show drop-abundance profiles (Figs.\ \ref{fig:abuncuthh46} and \ref{fig:abuncutiras2a} in the online appendix). CO is not expected to freeze out in the less massive and warmer envelope of DK~Cha, so the horizontal cuts merely show the effect of the photodissociation rate decreasing with distance to the cavity wall (Figs.\ \ref{fig:abunco} and \ref{fig:abuncutdkcha}).

\onlfig{10}{
\begin{figure*}[t]
\centering
\includegraphics[width=\hsize]{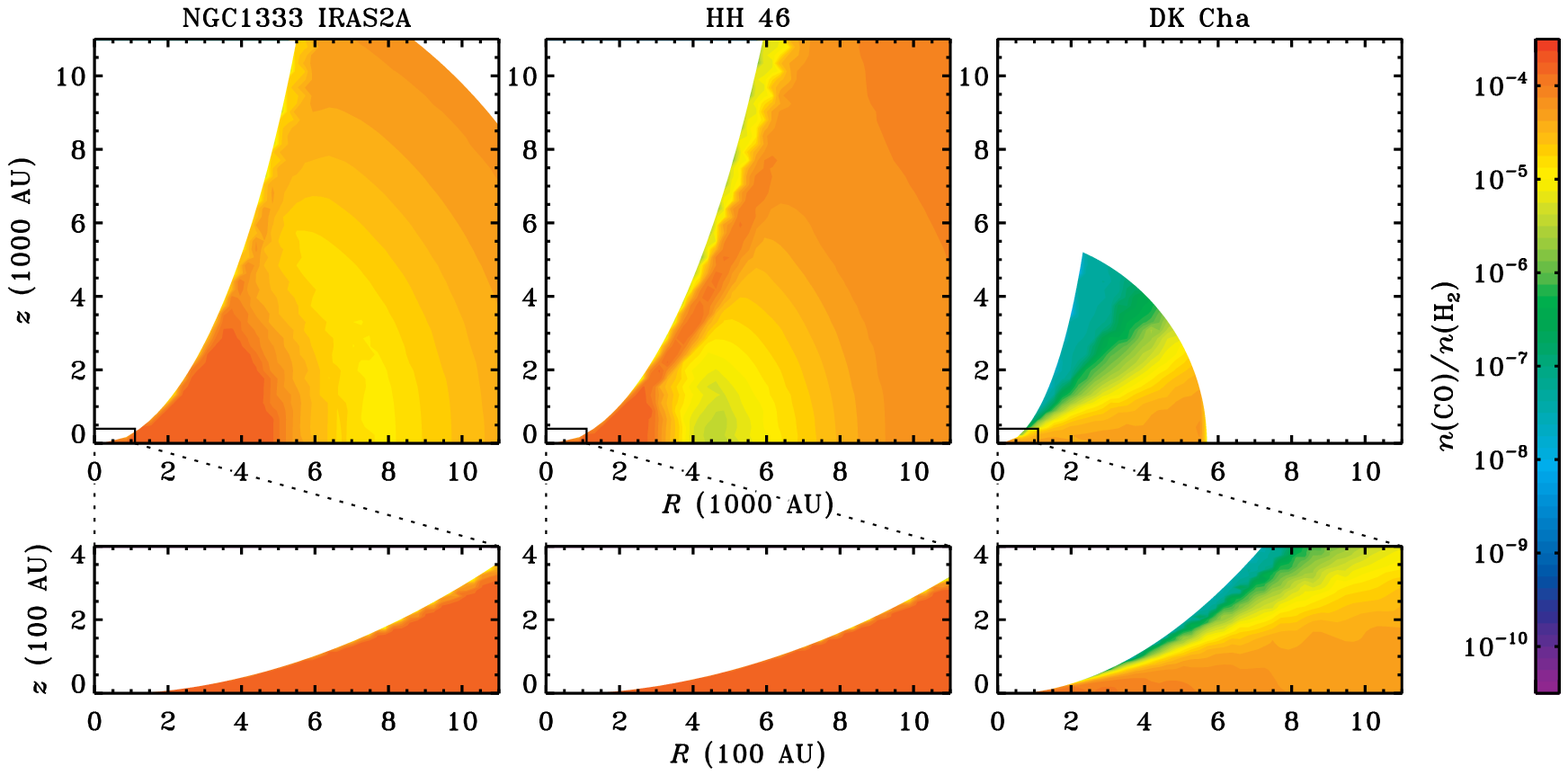}
\caption{Abundance profiles for CO for the model with outflow cavities and UV heating. Shocks are not included. The bottom row zooms in on the regions indicated in the top row.}
\label{fig:abunco}
\end{figure*}
}

The protostellar UV field also dissociates \w{}, making for a nearly uniform abundance of \ten{-10} along the wall of each of the three sources (\fig{cavabun}). Moving into the envelope, the increasing amount of extinction slows down the photodissociation process and the \w{} abundance increases to a few \ten{-8} in the two regions where it does not freeze out: the warm inner envelope and the low-density outer envelope (\fig{simpleabun}, bottom right). In between, freeze-out keeps the abundance limited to at most a few \ten{-9}.
As in the case of CO, the original 1D abundance profiles largely survive at the midplane (Figs.\ \ref{fig:abunh2o}--\ref{fig:abuncutdkcha} in the online appendix).

\onlfig{11}{
\begin{figure*}[t]
\centering
\includegraphics[width=\hsize]{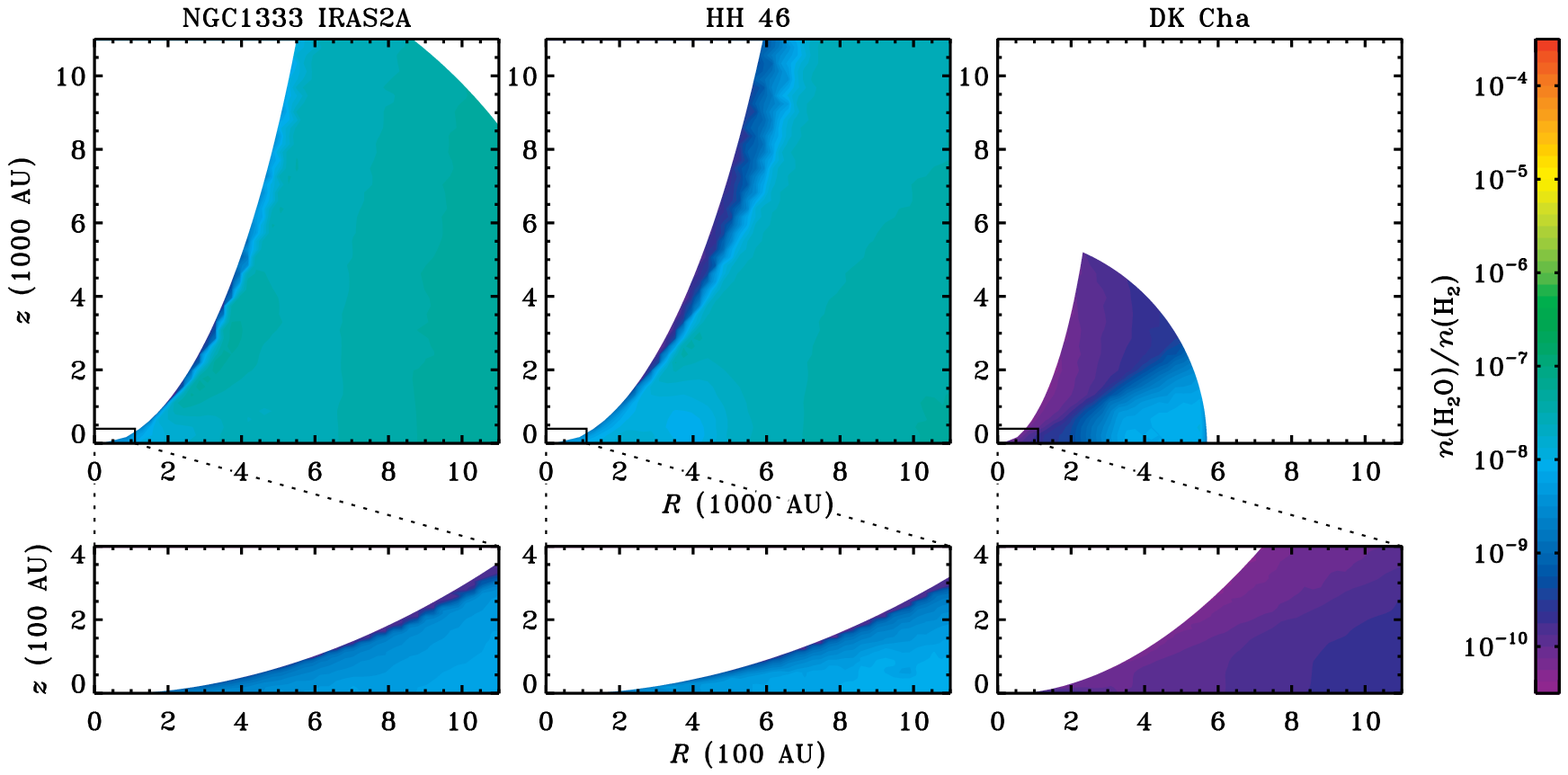}
\caption{As \fig{abunco}, but for \w\@.}
\label{fig:abunh2o}
\end{figure*}
}

\onlfig{12}{
\begin{figure*}[t]
\centering
\includegraphics[width=\hsize]{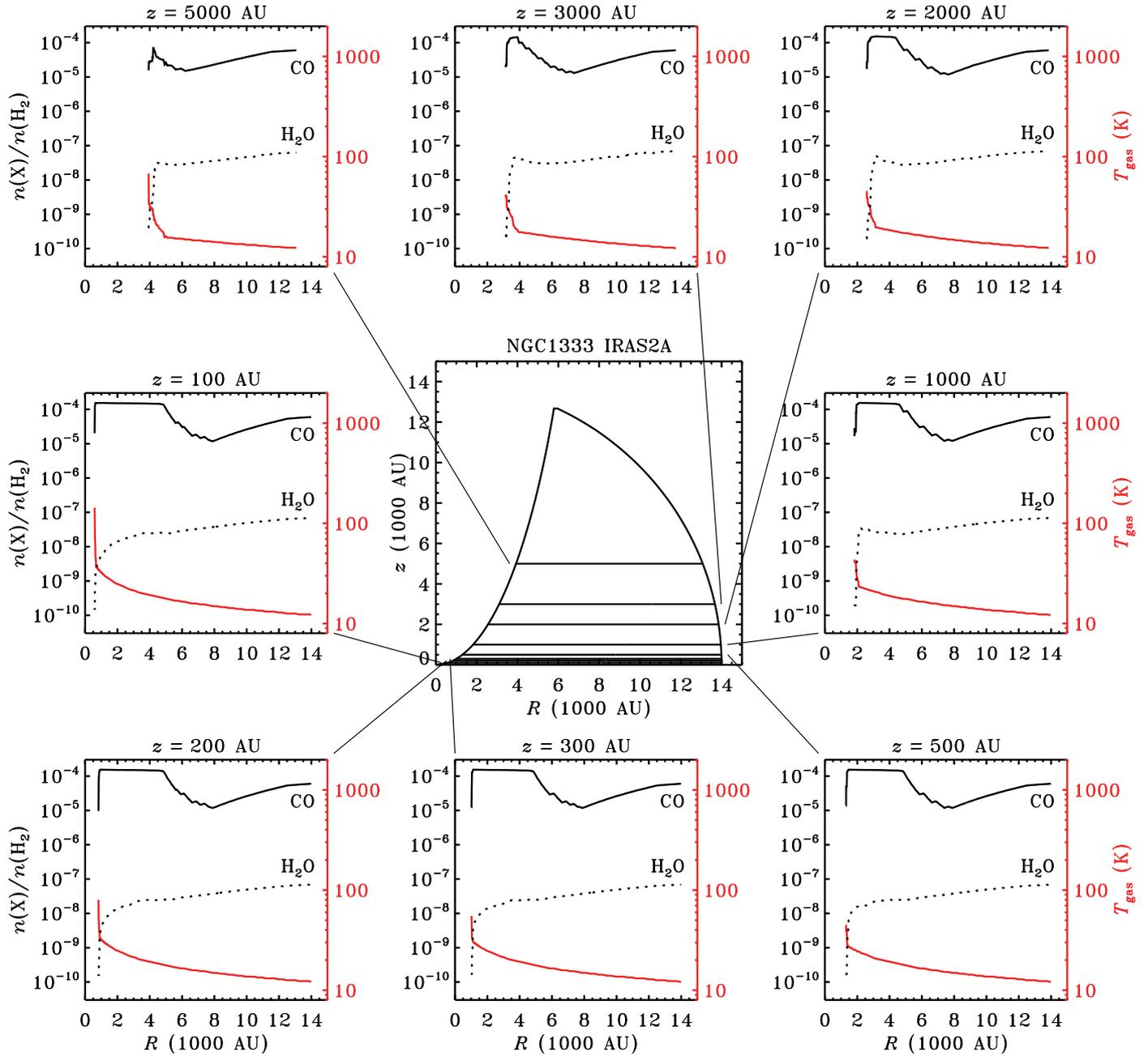}
\caption{Gas temperature (red) and abundances of CO and \w{} (solid and dotted black) in eight horizontal cuts through the envelope of NGC1333 IRAS2A.}
\label{fig:abuncuthh46}
\end{figure*}
}

\onlfig{13}{
\begin{figure*}[t]
\centering
\includegraphics[width=\hsize]{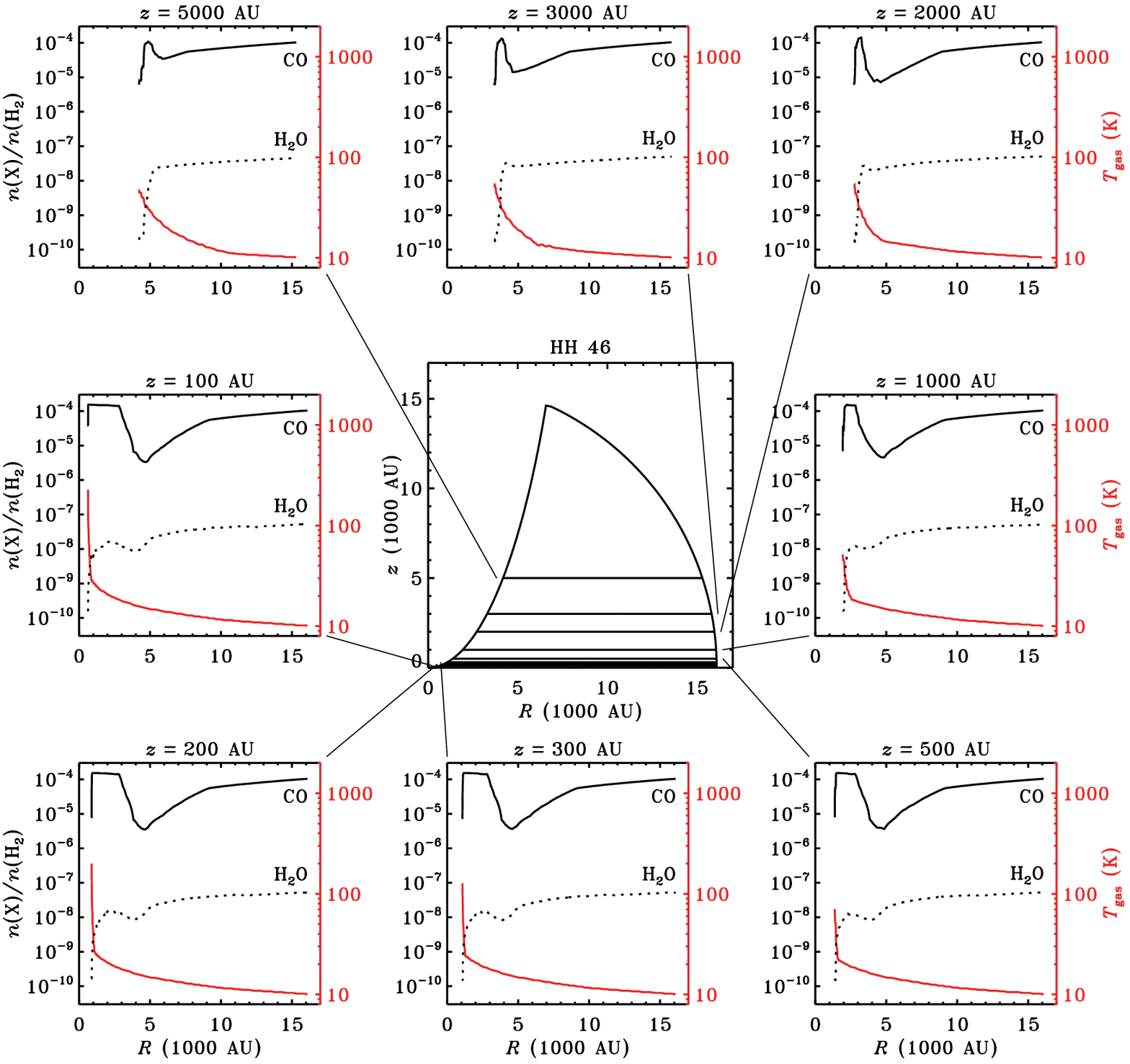}
\caption{Gas temperature (red) and abundances of CO and \w{} (solid and dotted black) in eight horizontal cuts through the envelope of HH~46.}
\label{fig:abuncutiras2a}
\end{figure*}
}

\onlfig{14}{
\begin{figure*}[t]
\centering
\includegraphics[width=\hsize]{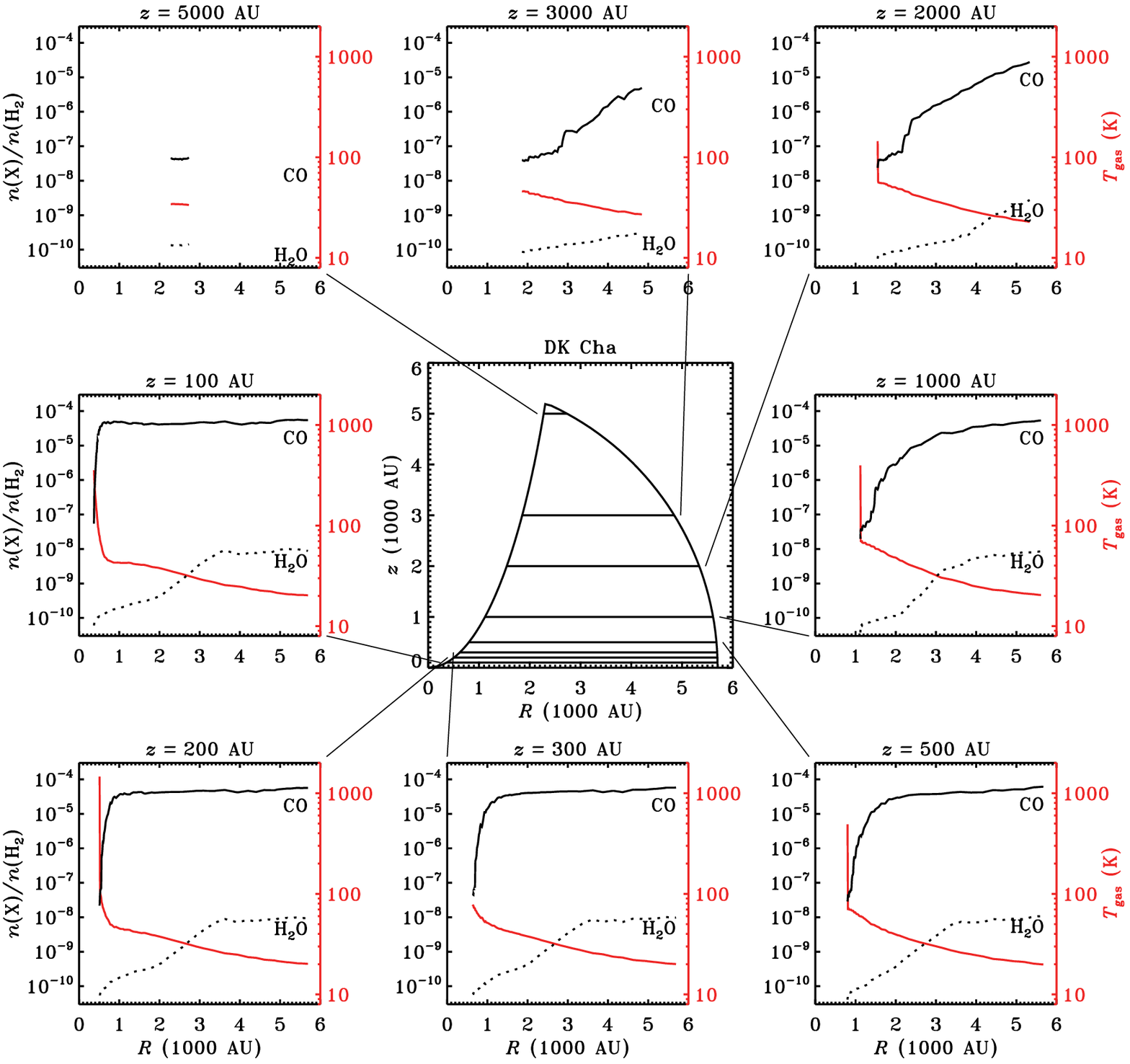}
\caption{Gas temperature (red) and abundances of CO and \w{} (solid and dotted black) in eight horizontal cuts through the envelope of DK~Cha.}
\label{fig:abuncutdkcha}
\end{figure*}
}

With the conditions in the passively heated gas and the UV-heated gas established, the final question is what happens in the small-scale shocks along the cavity walls. As shown in \fig{cool1}, the peak temperature in a 20 km \ps{} shock ranges from 1300 to 3800 K for pre-shock densities from \ten{4} to \ten{6.5} \pcc. In addition, the peak temperature depends roughly linearly on the shock velocity \citep[e.g.,][]{flower10a}. The higher temperatures help overcome the barriers on the ${\rm O}\ +\ \mhm\ \to\ {\rm OH}\ +\ {\rm H}$ and ${\rm OH}\ +\ \mhm\ \to\ {\rm H}_2{\rm O}\ +\ {\rm H}$ reactions \citep{bergin98a}, so the \w{} abundance in the shocked gas is expected to be orders of magnitude higher than the value of \ten{-10} calculated for the UV-heated gas (\fig{cavabun}).

The shock models of \citet{flower03a} predict a peak \w{} abundance of $\sim$\ten{-4} (\fig{cool1}), but that is in the absence of a UV field (Sect.\ \ref{subsec:cshock}). The photodissociation and reformation timescales can be estimated to get an idea of how much the \w{} abundance would change in an irradiated shock. At 1000 AU from the protostar, the UV flux is equivalent to $G_0\approx100$, which gives a photodissociation timescale of about 1 yr \citep{vandishoeck06a}. A reformation timescale of about 1 yr occurs for $T=1000$ K and $n(\mhm)=10^5$ \pcc{} \citep{bergin98a}. The density at 1000 AU is closer to \ten{6} \pcc{} (\fig{1denv}), so the reformation timescale is a factor of a few shorter than the photodissociation timescale. However, the two numbers are similar enough that in reality either process could dominate over the other, in particular when considering different points along the cavity wall. A new generation of shock models, incorporating UV radiation, is needed to solve this issue. Until then, the peak \w{} abundance of $\sim$\ten{-4} in \fig{cool1} should be considered an upper limit only.


\subsection{UV luminosities and shock velocities}
\label{subsec:chisq}
The model has two free parameters -- the UV luminosity and the shock velocity -- for which the best-fit values are obtained by comparing the computed CO line fluxes to the available \emph{Herschel} observations in a $\chi^2$ procedure. The lower-$J$ CO lines observed from the ground are not included in the fit. They contain contributions from the parent molecular cloud and from colder swept-up gas, neither of which is present in our model. The \w{} observations are also not used to constrain the model. The grid of UV luminosities runs from 0.01 to 1 $L_\odot$. The grid of shock velocities runs from 10 to 35 km \ps, approximately the critical velocity of C-type shocks for the relevant density regime (Sect.\ \ref{subsec:cshock}).

\begin{figure}[t]
\centering
\includegraphics[width=\hsize]{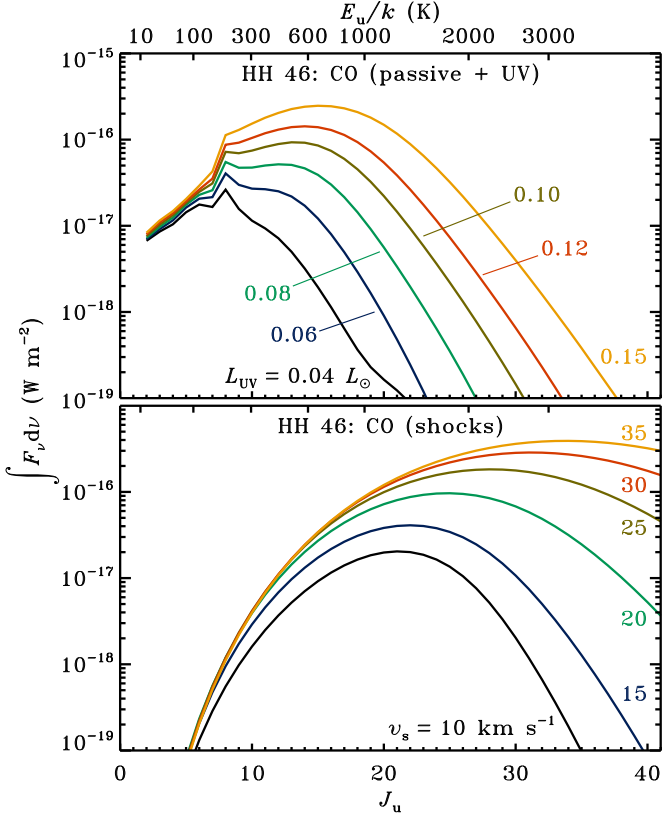}
\caption{\emph{Top:} CO fluxes computed for the passive and UV components in HH~46 together, for UV luminosities ranging from 0.06 to 0.15 $L_\odot$. \emph{Bottom:} CO fluxes computed for the shock component in HH~46, for shock velocities ranging from 10 to 35 km \ps.}
\label{fig:lvgrid}
\end{figure}

As an example of the dynamic range of the free parameters, \fig{lvgrid} shows the predicted line fluxes for HH~46 for a range of values. The top panel shows how the combined flux from the passively heated envelope and the UV-heated outflow cavity walls changes as function of UV luminosity. The bottom panel does so for the flux from the small-scale shocks as function of shock velocity. These figures are qualitatively the same for the other two sources.

\begin{figure}[t]
\centering
\includegraphics[width=\hsize]{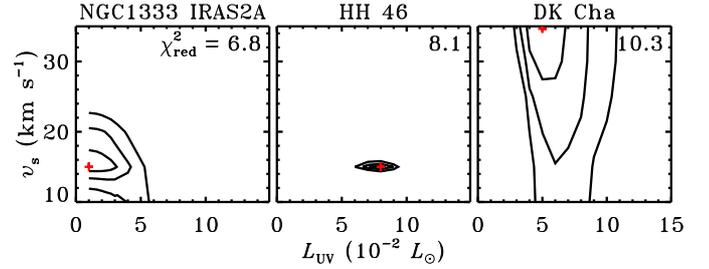}
\caption{Reduced $\chi^2$ contours for the UV luminosity and the shock velocity in each source. Red crosses mark the minima. Contours are at the 1, 2 and $3\sigma$ levels.}
\label{fig:chisq}
\end{figure}

The best-fit UV luminosities and shock velocities are tabulated in \tb{modpar} and shown in \fig{chisq}. Overplotted in the figure are the 1, 2 and $3\sigma$ confidence intervals. Both parameters are well constrained in HH~46. In IRAS2A, the UV luminosity is only constrained from the top. In DK~Cha, the shock velocity is unconstrained at the 3$\sigma$ level and only constrained from below at the 2$\sigma$ level. Extending the grid of luminosities to lower values does not resolve the issue for IRAS2A, because the gas cannot cool below the dust temperature. The $\chi^2$ analysis suggests that the CO fluxes observed with \emph{Herschel} are powered primarily by shocks in IRAS2A, by UV-heating in DK~Cha, and by a combination of both in HH~46. This result is discussed in more detail in the next section.

\subsection{Line fluxes}
\label{subsec:lflux}


\begin{figure*}[t]
\centering
\includegraphics[width=\hsize]{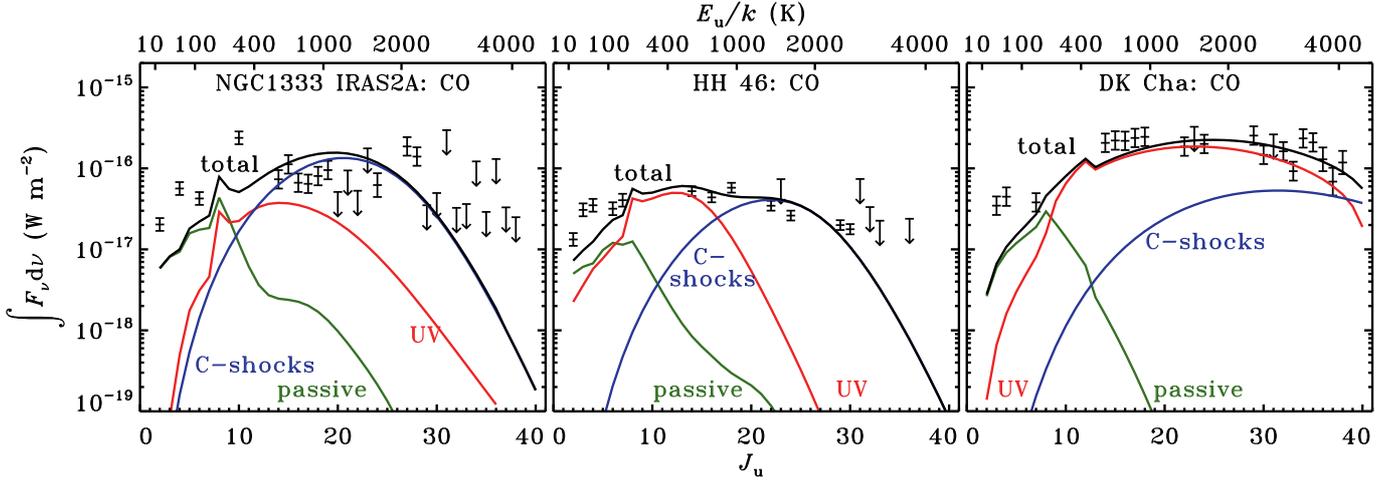}
\caption{CO line fluxes observed with the JCMT, APEX and \emph{Herschel}. Overplotted are the fluxes from the three model components (green, red and blue) and their sums (black). Error bars (1$\sigma$) reflect the 10--30\% calibration uncertainty of \emph{Herschel} and the 20--30\% calibration uncertainty of the JCMT and APEX\@. Upper limits are at the $3\sigma$ level.}
\label{fig:coladder}
\end{figure*}

\subsubsection{Carbon monoxide}
\label{subsubsec:coflux}
With the physical and chemical structure from Sect.\ \ref{subsec:physchem} and the best-fit UV luminosities and shock velocities from Sect.\ \ref{subsec:chisq} (taking $L\q{UV}=0.01$ $L_\odot$ for IRAS2A and $\varv\q{s}=35$ km \ps{} for DK~Cha), the model produces the integrated CO line fluxes shown in \fig{coladder}. Also plotted are the integrated line fluxes observed from the ground and with \emph{Herschel}. The only systematic discrepancy between model and observations is an underproduction of the 2--1, 3--2 and 4--3 lines. The same discrepancy was seen by \citet{vankempen09b} in their purely spherical model of HH~46. The excess flux probably originates in cold gas in the parent molecular cloud or in cold swept-up outflow gas, neither of which is included in our model. The low-$J$ lines of the less abundant C$^{18}$O isotopologue are not expected to have any significant cloud or outflow contribution \citep[e.g.,][]{jorgensen02a,yildiz10a}, and indeed the model reproduces the available C$^{18}$O data to within 50\%. The mismatches with some of the individual \emph{Herschel} data points in \fig{coladder} do not appear to follow any systematic trend. They are likely due to uncertainties in the model fluxes and minor factors such as pointing offsets and calibration uncertainties that may be larger than the estimated 10--30\%.

The coloured curves in \fig{coladder} show the predicted contribution from the passively heated envelope (green), the UV-heated cavity walls (red) and the small-scale shocks (blue) to the total CO emission (black). Each source requires a combination of UV-heated gas and small-scale shocks to reproduce the \emph{Herschel} data, although there are clear qualitative differences from source to source. IRAS2A is the youngest of the three sources (deeply embedded Stage 0) and its CO ladder appears to be dominated by shock-powered emission. HH~46 is more evolved (Stage I) and shows a 50/50 split between UV-powered and shock-powered emission. DK~Cha is still more evolved (in transition from Stage I to II) and its small-scale shocks are responsible for only a small fraction of the overall CO emission. The decreasing contribution of the shock component is consistent with the general trend of the outflow force being weaker in more evolved protostars \citep{bontemps96a}. The increasing contribution of the UV component is powered by the decreasing densities in the envelope, allowing the gas to become hotter (Figs.\ \ref{fig:1denv} and \ref{fig:cavtemp}; see also Sect.\ \ref{subsec:tgas}). Although the sample size is small, the observed data are entirely in keeping with these physically motivated trends.
Application of the model to a larger number of sources, such as the full WISH sample, will help to obtain a firmer conclusion. Furthermore, a full parameter study is planned to assess systematically how the CO emission depends on properties such as the envelope mass and the UV luminosity.

\begin{table}
\caption{Absolute and relative luminosities of CO and \w{} for the three physical components in each source.}
\label{tb:decomp}
\centering
\begin{tabular}{l c@{\ }c c@{\ }c c@{\ }c}
\hline\hline
\cellcent{Component} & \multicolumn{2}{c}{IRAS2A} & \multicolumn{2}{c}{HH~46} & \multicolumn{2}{c}{DK~Cha} \\
\cellcent{for CO}    & \ten{-3} $L_\odot$ & \% & \ten{-3} $L_\odot$ & \% & \ten{-3} $L_\odot$ & \% \\
\hline
passive              & 0.3 &   7 & 0.5 &   7 & 0.1 &   2 \\
UV                   & 0.8 &  20 & 3.1 &  45 & 4.0 &  78 \\
C-shocks             & 3.0 &  73 & 3.3 &  48 & 1.0 &  20 \\
\hline
total                & 4.1 & 100 & 6.9 & 100 & 5.1 & 100 \\
\hline
\\
\hline\hline
\cellcent{Component} & \multicolumn{2}{c}{IRAS2A} & \multicolumn{2}{c}{HH~46} & \multicolumn{2}{c}{DK~Cha} \\
\cellcent{for \w}    & \ten{-3} $L_\odot$ & \% & \ten{-3} $L_\odot$ & \% & \ten{-3} $L_\odot$ & \% \\
\hline
passive              & 0.02    & 0.02    & 0.53 & 0.38 & 0.05    & 0.17 \\
UV                   & 0.001   & 0.001   & 0.04 & 0.03 & $3(-5)$ & $1(-4)$ \\
C-shocks             & 100     & $>$99.9 & 140  & 99.6 & 30      & 99.8 \\
\hline
total                & 100     & 100     & 140  & 100  & 30      & 100 \\
\hline
\end{tabular}
\tablefoot{The tabulated values are for the adopted source inclinations of 90, 53 and 0$^\circ$ (\tb{modpar}). They do not represent the total line cooling budget, which would require integrating over all inclinations from edge-on to pole-on.}
\end{table}

The contribution in each curve can be quantified by summing the line fluxes from $\jup=2$ to 40 and correcting them for distance to give the overall CO luminosities for the adopted inclinations (\tb{decomp}).
The CO line fluxes can also be plotted in a rotational diagram to investigate excitation conditions. \tb{trot} lists the rotational temperatures derived by fitting straight lines to the passive curve from $\eup=10$ to 100 K (up to $J=5$--4), to the UV curve from 100 to 1000 K (up to $J=18$--17), and to the shock curve from 1000 to 4000 K\@. The error margins reflect the stochastic $1\sigma$ uncertainty on each fit, but do not account for any uncertainties in the model results.

\begin{table}
\caption{Rotational temperatures for CO and \w{} in the three physical components in each source.}
\label{tb:trot}
\centering
\begin{tabular}{lccc}
\hline\hline
\cellcent{Component} & IRAS2A & HH~46 & DK~Cha \\
\cellcent{for CO}    & $T\q{rot}$ (K) & $T\q{rot}$ (K) & $T\q{rot}$ (K) \\
\hline
passive              & $31\pm5$   & $33\pm9$   & $54\pm5$ \\
UV                   & $240\pm30$ & $160\pm20$ & $410\pm100$ \\
C-shocks             & $380\pm20$ & $400\pm20$ & $1600\pm50$ \\
\hline
\\
\hline\hline
\cellcent{Component} & IRAS2A & HH~46 & DK~Cha \\
\cellcent{for \w}    & $T\q{rot}$ (K) & $T\q{rot}$ (K) & $T\q{rot}$ (K) \\
\hline
passive              & $72\pm1$   & $130\pm5$   & $175\pm10$ \\
UV                   & $30\pm1$   & $69\pm1$    & $114\pm1$ \\
C-shocks             & $410\pm80$ & $540\pm170$ & $750\pm400$ \\
\hline
\end{tabular}
\end{table}

The CO ladder for DK~Cha in \fig{coladder} implies that the six CO lines seen with ISO, from $J=14$--13 to 19--18 \citep{giannini99a}, originate in the UV-heated gas in the outflow cavity walls. Based on a passively heated spherical envelope model, \citet{vankempen06a} concluded that the ISO lines trace material on larger spatial scales than the material responsible for the 7--6 and lower lines observed with APEX\@. With the more complex source geometry in our model, such a requirement no longer exists. In fact, the temperature gradient along the cavity wall (\fig{cavtemp}), together with the pole-on inclination, suggests the high-$J$ lines seen with ISO originate on smaller spatial scales than the lower-$J$ APEX lines. A more detailed analysis of the spatial extent of the PACS data, using all $5\times5$ spaxels, can help to solve the puzzle, but this is beyond the scope of the current work.


\begin{figure*}[t]
\centering
\includegraphics[width=\hsize]{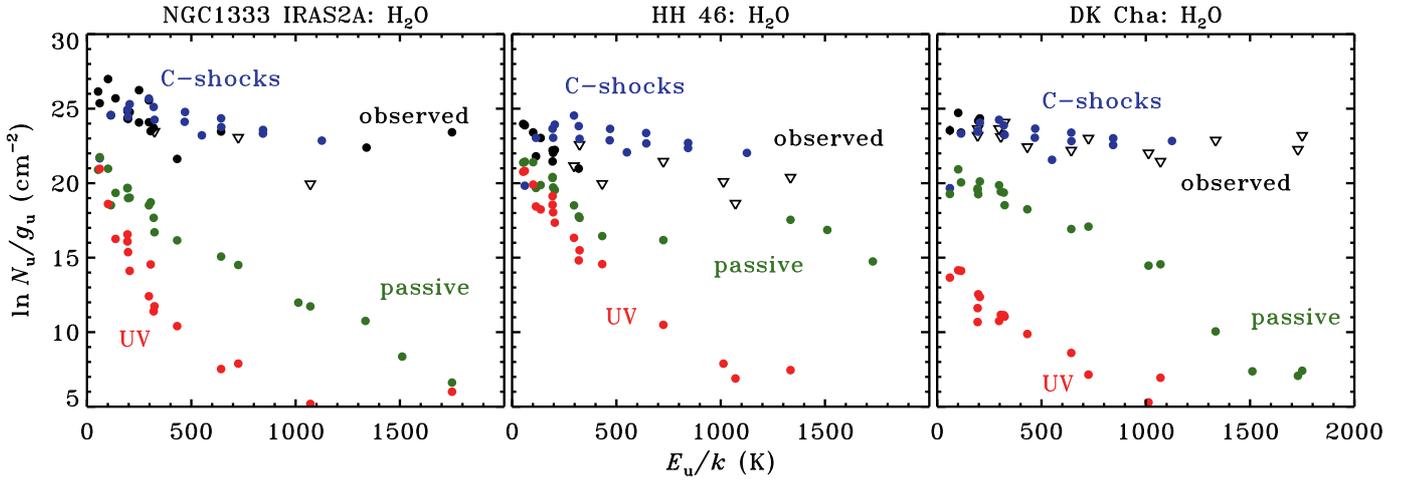}
\caption{Rotational diagrams for water. Black: \emph{Herschel} detections (circles) and $3\sigma$ upper limits (triangles). Green, red and blue: model predictions for the passive, UV and shock components.}
\label{fig:h2oladder}
\end{figure*}

\subsubsection{Water}
\label{subsubsec:wflux}
\figg{h2oladder} shows the observations and model predictions for \w\@. The complicated rotational structure of \w{} does not allow for plots in the format of the CO ladders in \fig{coladder}, so the results are shown as rotational diagrams instead. The individual line fluxes can be summed again to obtain the total line luminosities. However, the models only calculate fluxes for a subset of all possible rotational lines. Fluxes for the missing lines in the HIFI and PACS wavelength ranges (up to $\eup/k=2000$ K) are calculated by fitting a single rotational temperature to the available lines in each model component (\tb{trot}). The level populations of \w{} are not in LTE, so this method may produce large errors in the flux of a given individual line. However, because those errors tend to cancel when summing over all possible lines, assuming a single temperature is appropriate for calculating the total line luminosities \citep[see also][]{goldsmith99a}. The resulting values are tabulated in \tb{decomp}.

Unlike CO, the \w{} emission observed with \emph{Herschel} seems to be dominated ($>$99\%) by shock-heating in all three sources. This is consistent with the broad \w{} line profiles seen with HIFI in IRAS2A and other embedded YSOs (\citealt{kristensen10b}, subm.). Our shock models actually overproduce many of the detections and upper limits by up to a factor of 10, in particular in HH~46 (\fig{h2oladder}). The likely cause is that the \w{} abundances in our shock models were calculated without taking the protostellar UV field into account. Based on the photodissociation and reformation timescales estimated in Sect.\ \ref{subsec:physchem}, the \w{} abundance can easily be lowered by a factor of 2 to 10 from the peak value of \ten{-4} shown in \fig{cool1}. At the same time, the abundance of OH would go up, so our shock models are currently expected to underproduce the OH line intensities. However, as argued by \citet{vankempen10a} and \citet{wampfler10a}, the OH lines observed with \emph{Herschel} likely originate in a different physical component than the \w{} lines (a dissociative J-type shock instead of a non-dissociative C-type shock). It is therefore not possible to test whether the OH line intensities predicted for the small-scale shocks are indeed lower than the actual OH emission from that component. A potential problem with lowering the \w{} abundances through photodissociation is that some CO in the shock may be dissociated as well. If that happens, a higher shock velocity is required to fit the CO observations, which in turn would raise the model \w{} fluxes back to their original values of up to an order of magnitude too high. An independent constraint on the CO and \w{} abundances in UV-irradiated shocks is needed to fully resolve this issue.

Another difference between CO and \w{} is seen in the relative contributions of the passive and UV components. For \w, both are weak compared to the shock-powered emission, but the passive component is always the brighter of the two (\tb{decomp}). The weak emission from the UV-heated gas in the outflow cavity walls is due to the very low \w{} abundance (Figs.\ \ref{fig:cavabun} and \ref{fig:simpleabun}), which in turn is due to the wide range of photon energies over which \w{} can be dissociated. CO can only be dissociated by photons of at least 11 eV, so its abundance in the cavity walls remains higher and the UV-heated gas contributes a larger fraction of the overall CO emission.


\section{Discussion}
\label{sec:disc}


\subsection{Gas temperature}
\label{subsec:tgas}
The largest unknown in our model is the gas temperature due to UV heating of the outflow cavity walls. Calculating the gas and dust temperatures for a given density and UV flux is a standard problem in models of photon-dominated regions \citep[PDRs; e.g.,][]{tielens85a,hollenbach97a}. However, the PDR community has not yet converged on a unique solution: in a benchmark test of ten different PDR codes, the temperature solutions varied by up to an order of magnitude \citep{rollig07a}. As shown in Sect.\ \ref{subsec:lflux}, the temperature grid from \citet[hereafter K99]{kaufman99a}, along with the $\exp(-0.6\av)$ depth dependence derived from the benchmark test, works well to reproduce the CO observations. However, given the complicated nature of our model, this does not necessarily mean that the \knn{} grid gives the correct temperatures. For example, if the grid systematically overpredicts the temperature, it may still give a good match to the data by decreasing the UV luminosity. Other temperature grids may also be able to match the data. The goal of this section is two-fold: (1) to show that, at least on a qualitative level, the \knn{} grid follows the density and UV flux dependence expected from basic heating and cooling physics for the relevant range of conditions; and (2) to see how robust our results are to changes in the adopted temperature grid.

\begin{figure}[t]
\centering
\includegraphics[width=\hsize]{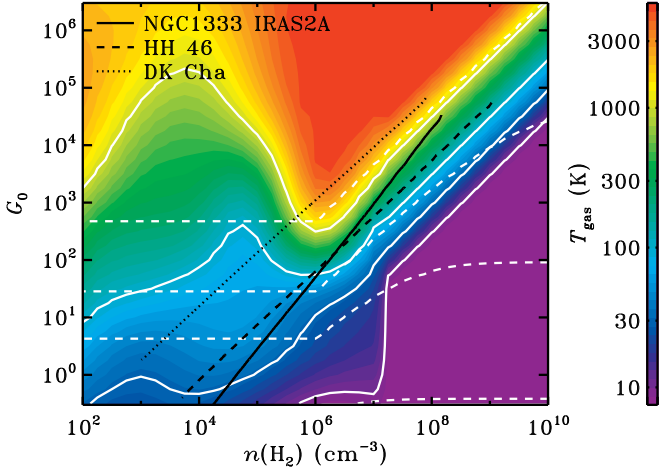}
\caption{PDR surface temperature grid (colour scale and solid white contours) from \citet{kaufman99a}, extrapolated as $T\q{gas} \propto n^{-1}$ for $n(\mhm)>10^7$ \pcc. The white contours are drawn at 10, 30, 100 and 1000 K\@. Overplotted as dashed white contours (same values) is the approximate temperature grid based on Eqs.\ (\ref{eq:ggcool}) and (\ref{eq:peheat}). The black lines trace the conditions along the cavity walls of the three sources.}
\label{fig:k99grid}
\end{figure}

The \knn{} grid is reproduced in \fig{k99grid}. In its original form, it only covers \mh{} densities up to \ten{7} \pcc. The densities in our model go up to \ten{9} \pcc, so the grid is extrapolated as $T\q{gas}\propto1/n(\mhm)$. The $n^{-1}$ dependence comes from the approximate functional forms of the heating and cooling mechanisms expected to dominate at high densities: photoelectric heating \citep[$\propto n$;][]{bakes94a} and gas-grain collisional cooling \citep[$\propto n^2$;][]{hollenbach89a}. The density dependence of the temperature in the original \knn{} grid is almost exactly $n^{-1}$ between \ten{6} and \ten{7} \pcc.

Overplotted as black lines in \fig{k99grid} are the gas density and incident UV flux along the cavity walls of the three sources. The density and UV flux both fall off approximately as $r^{-2}$, so they scale linearly with each other. At densities above $\sim$\ten{6} \pcc, the gas temperature is expected to scale with the inverse of the density (see above) and linearly with the UV flux \citep{bakes94a}, and should therefore remain constant along the inner part of the wall. This is indeed seen in \fig{cavtemp}: $T\q{gas}$ does not change much over the range of radii corresponding to densities of more than \ten{6} \pcc. Photoelectric heating likely remains dominant at lower densities, but atomic fine-structure lines take over from gas-grain collisions as the dominant cooling mechanism. Their cooling rate depends linearly on the density as long as the density exceeds the critical density of \scit{3}{3} \pcc{} for the [\ion{C}{ii}] 158 \micron{} line, so the temperature is expected to remain roughly constant with density \citep[\knn;][]{meijerink07a}. Because the $G_0$ dependence remains, the temperature along the cavity wall should decrease as $r^{-1}$ for $n(\mhm)<10^6$ \pcc. Again, this is the approximate behaviour seen in \fig{cavtemp}. The dependencies change again for densities below \scit{3}{3} \pcc{}, but such low densities do not appear in our model.

The location of the black lines in \fig{k99grid} shows why the UV-powered CO emission looks qualitatively different in DK~Cha than in IRAS2A and HH~46. The lower envelope densities in the former turn the conditions along the cavity wall into a classic PDR, with densities of \ten{6}--\ten{7} \pcc{} and UV fluxes of \ten{3}--\ten{4} times the mean interstellar flux. The gas therefore gets much hotter and CO is excited to higher rotational levels. The steep temperature gradient in this part of \fig{k99grid} also means that the CO emission is rather sensitive to details in the SED fitting with DUSTY\@. A factor of two difference in density leads to a similar change in the gas temperature, which can affect the highest-$J$ CO lines by a few orders of magnitude. \figg{coladder} tentatively shows an evolutionary trend of the CO emission in more evolved sources being more strongly UV-powered. While the application of a physical model is necessary to quantify exactly how much of the total CO ladder in a given source is due to UV heating, the observations are consistent with the main point that with decreasing density (assuming sufficient UV), sources becomes more PDR-like. This is what one would naturally expect in the transition from Stage 0 to Stages I and II objects as presented in this paper.

Of equal importance to the CO emission is the choice of the PDR temperature grid. An easy way to get a rough idea of how changes in the adopted grid would affect the CO line fluxes is to parameterise the gas temperature based on the dominant heating and cooling mechanisms. Two equations suffice to do so. The first one approximates the gas-grain collisional cooling rate \citep{hollenbach89a}:
\begin{align}
\label{eq:ggcool}
\Lambda\q{g-g} =\ & (\scim{1}{-31}\,{\rm erg}\,{\rm cm}^{-3}\,{\rm s}^{-1}) n\q{H}^2 \sqrt{\frac{T\q{gas}}{1000\,{\rm K}}} \sqrt{\frac{100\,\text{\AA}}{a\q{min}}} \nonumber \\
                   & \left[1-0.8 \exp\left(-\frac{75\,{\rm K}}{T\q{gas}}\right)\right] (T\q{gas}-T\q{dust})\,,
\end{align}
where $n\q{H}$ is the total hydrogen nucleus density (equal to $2n(\mhm)$ in our model for all practical purposes) and $a\q{min}$ is the minimum grain size (set to 10 \AA). The second equation is for the photoelectric heating rate \citep{bakes94a}:
\begin{equation}
\label{eq:peheat}
\Gamma\q{pe} = (\scim{1}{-24}\,{\rm erg}\,{\rm cm}^{-3}\,{\rm s}^{-1}) \epsilon G_0 n\q{H}\,.
\end{equation}
The photoelectric heating efficiency, $\epsilon$, is treated as a free parameter for the purpose of this simple approximation. As such, it hides any uncertainties in the grain charge and in the numerical prefactors in the two equations \citep[see, e.g.,][]{young04a}. Given a dust temperature from RADMC (or, alternatively, from the analytical expression of \citealt{hollenbach91a}), the gas temperature at $n(\mhm)>10^6$ \pcc{} is obtained by solving for $\Lambda\q{g-g}=\Gamma\q{pe}$. At lower densities, the dependence of $T\q{gas}$ on $n(\mhm)$ disappears to zeroth order, and we set $T\q{gas}[G_0,n(\mhm)<10^6\ {\rm cm}^3]=T\q{gas}[G_0,10^6\ {\rm cm}^3]$.

\figg{k99grid} shows the 10, 30, 100 and 1000 K contours for $\epsilon=0.2$ as dashed white lines. The approximate grid reproduces the \knn{} grid to within a factor of 2 for the full range of densities and UV fluxes encountered along the three cavity walls. The approximation breaks down for $G_0/n(\mhm)<10^{-4}$ (high density, low UV) and $G_0/n(\mhm)>10^{-3}$ (low density, high UV), and cannot be used in either of those regimes. The value of 0.2 required for a good match with \knn{} exceeds the maximum value of 0.05 allowed by \citet{bakes94a}. However, the approximate temperature grid is purely meant as an easy tool to study the effect of different temperature grids, so this factor of 4 should not be considered physically important. The key point is that the \knn{} grid and the approximate grid show the same qualitative trends for the relevant range of conditions, which means the \knn{} grid follows the behaviour expected from the dominant heating and cooling mechanisms.

\begin{figure}[t]
\centering
\includegraphics[width=\hsize]{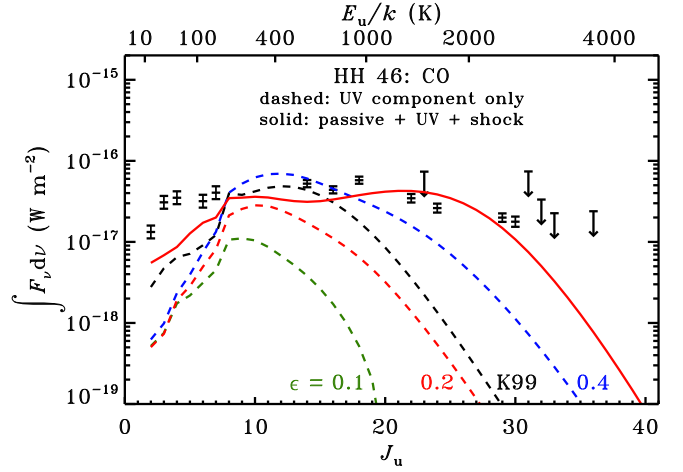}
\caption{CO ladder for HH~46. The data points are as in \fig{coladder}. The dashed black curve is the model flux for the UV component for temperatures from the \citet{kaufman99a} grid. The dashed coloured curves are the same, but for the temperature approximation from Eqs.\ (\ref{eq:ggcool}) and (\ref{eq:peheat}), using different heating efficiencies. The solid red curve combines the dashed red curve with the passive and shock curves from \fig{coladder}.}
\label{fig:tgasco}
\end{figure}

With $\epsilon=0.2$, the approximate grid does a good job of reproducing the CO line emission for the UV component. Shown in \fig{tgasco} are the integrated line fluxes from the UV-heated gas in HH~46, using either the \knn{} grid (dashed black curve, identical to the red curve from \fig{coladder}) or the approximate grid with three different values of $\epsilon$ (dashed coloured curves). The $\epsilon=0.2$ curve lies within 30\% of the \knn{} curve up to $\jup=20$, at which point shocks take over as the dominant producer of CO emission. When adding the passive and shock components to the $\epsilon=0.2$ curve (producing the solid red curve), the overall agreement with the data is as good as in \fig{coladder}. Taking a lower or higher photoelectric heating efficiency would under- or overproduce the $J=14$--13, 16--15 and 18--17 lines.

The photoelectric heating rate scales linearly with $\epsilon$ and $G_0$ (\eq{peheat}), so the three values of $\epsilon$ in \fig{tgasco} can be brought to give the same CO fluxes by changing the UV luminosity. Herein lies the degeneracy alluded to at the beginning of this section and in Sect.\ \ref{subsec:chisq}: as long as two temperature grids show the same qualitative trends, both can reproduce the CO observations, but they have different best-fit UV luminosities. This degeneracy can only be lifted by calibrating the temperature grids against a source whose incident UV flux can be constrained independently in the same part of $n(\mhm)$--$G_0$ parameter space.


\subsection{Caveats in the shock models}
\label{subsec:vshock}
The shock modelling relies on three assumptions: (1) the shocks move along the cavity walls, (2) the effective beam-filling factor can be obtained from accurately estimating the shock width, and (3) the protostellar UV field does not affect the \w{} abundance. In order to test where the emission is coming from in this scenario -- extended lower-density gas with a large beam-filling factor or higher-density gas with a small beam-filling factor -- fluxes from the standard shock model are compared to fluxes from models without any contribution from gas with densities of \ten{4} \pcc{} and lower, \ten{5} \pcc{} and lower, or \ten{6} \pcc{} and lower. The shock velocity is kept at 20 km \ps{} for this experiment. For the case of HH~46, gas at densities below \ten{4} \pcc{} is found to contribute less than \ten{-4}\%, and gas at densities between \ten{4} and \ten{5} \pcc{} only 0.3\%. About 75\% of the overall flux comes from the segment with a pre-shock density of \ten{6.5} \pcc. The other two sources show almost the same numbers, indicating that the shocked gas observed by PACS is dominated by emission from the densest parts of the YSO.

If the density is assumed to be fixed at more than \ten{6} \pcc, the shock velocity alone determines the absolute line fluxes. This was already seen in the bottom panel of \fig{lvgrid}, where the basic shape of the CO ladder is nearly independent of velocity. Hence, the velocity determines the intensity of the CO emission, and the pre-shock density determines the shape of the ladder.
Comparing the fluxes amongst the three sources, the differences are dominated by their respective distances.
This confirms that the actual geometry only plays a small role with respect to the predicted CO emission, and that the bulk of emission is created in very dense gas contained within the central PACS spaxel.

If the effective beam-filling factor is treated as a free parameter, it becomes degenerate with the shock velocity. A smaller beam-filling factor would require a higher shock velocity to produce the same line fluxes, and vice versa. However, for a given source geometry, such as the ellipsoid outflow cavity in our model, the beam-filling factor follows from the cooling lengths and is not truly a free parameter. The only way to constrain either the geometry or the beam-filling factor would be to have an independent handle on the pre-shock density and the shock velocity. Our method therefore gives a unique solution for the given shock geometry, but other geometries cannot be excluded.


\section{Conclusions}
\label{sec:conc}
The \emph{Herschel Space Observatory} offers a new probe into the hot gas present in embedded low-mass young stellar objects (YSOs) through emission lines of highly rotationally excited CO (up to 4000 K above the ground state) and \w{} (up to 600 K). This paper presents the first model to reproduce these observations quantitatively. The model starts with a passively heated spherical envelope, which fits the lowest-$J$ CO isotopologue lines observed from the ground, but falls short of the $\chm{^{12}}$CO \emph{Herschel} data by several orders of magnitude. Two additional model components are therefore invoked: (1) gas in the walls of a bipolar outflow cavity, exposed to the protostellar UV field, and (2) small-scale C-type shocks along the cavity walls.

The main conclusion is that, within the context of the model, the far-infrared CO rotational line emission in the low-mass YSOs NGC1333 IRAS2A, HH~46 and DK~Cha can only be reproduced with a combination of UV-heated gas and C-type shocks.
The three sources show a tentative evolutionary trend: the CO emission appears to be dominated by shock-heating in the youngest source (IRAS2A), by UV-heating in the oldest source (DK~Cha), and by a 50/50 combination in the source of intermediate age (HH~46). This trend is powered by the envelope density decreasing as a protostar evolves from Stage 0 to Stages I and II.

The \w{} lines are dominated by shocks in all three sources, with less than 1\% of the total \w{} line luminosity originating in UV- and passively heated gas. This is a direct result of the different abundances in the three components. Freeze-out keeps the \w{} abundance below \ten{-8} in the cold outer envelope. The dust grains in the UV-heated layer are warm enough for \w{} to evaporate, but it is dissociated by the UV field once it does. Efficient reformation at high temperatures boosts the abundance to \ten{-5}--\ten{-4} in the C-type shocks, so they are responsible for most of the \w{} emission. CO has a lower evaporation temperature (20 versus 100 K), so its abundance is orders of magnitude higher than that of \w{} in the passively heated and the UV-heated gas, allowing these two components to contribute more strongly to the overall CO emission.

The main uncertainty in the model lies in the calculation of the gas temperature as function of the incident UV flux. This is a well-known problem in photon-dominated regions (PDRs), and many solutions are available in the literature. However, they disagree by up to an order of magnitude and some do not even agree qualitatively on how the gas temperature depends on the density or the UV flux. The sensitivity of our model results to the treatment of the gas temperature is tested with an approximate analytical description of the expected dominant heating and cooling mechanisms. This exercise shows that the quantitative contributions of the UV-heated gas and the C-type shocks to the hot gas emission observed by \emph{Herschel} are uncertain by at least a factor of 2, but the qualitative conclusion that both components are needed to explain the observations is robust.


\begin{acknowledgements}
The authors are grateful to the WISH and DIGIT teams -- in particular Neal Evans, Jes J{\o}rgensen and Ted Bergin -- for stimulating discussions and scientific support. Constructive comments from the anonymous referee are kindly acknowledged.
Astrochemistry in Leiden is supported by the Netherlands Research School for Astronomy (NOVA), by a Spinoza grant and grant 614.001.008 from the Netherlands Organisation for Scientific Research (NWO), and by the European Community's Seventh Framework Programme FP7/2007--2013 under grant agreement 238258 (LASSIE).
Partial support for this work was provided by NASA through RSA awards No.\ 1371476 and 1416374, by the NSF through grant AST-1008800, and by JPL through Contract 1358118.
HIFI has been designed and built by a consortium of institutes and university departments from across Europe, Canada and the United States under the leadership of SRON Netherlands Institute for Space Research, Groningen, the Netherlands, and with major contributions from Germany, France and the USA\@. Consortium members are: Canada: CSA, U.\ Waterloo; France: CESR, LAB, LERMA, IRAM; Germany: KOSMA, MPIfR, MPS; Ireland: NUI Maynooth; Italy: ASI, IFSI-INAF, Osservatorio Astrofisico di Arcetri-INAF; Netherlands: SRON, TUD; Poland: CAMK, CBK; Spain: Observatorio Astron{\'o}mico Nacional (IGN), Centro de Astrobiolog{\'\i}a (CSIC-INTA); Sweden: Chalmers University of Technology (MC2, RSS \& GARD), Onsala Space Observatory, Swedish National Space Board, Stockholm University (Stockholm Observatory); Switzerland: ETH Zurich, FHNW; USA: Caltech, JPL, NHSC\@.
PACS has been developed by a consortium of institutes led by MPE (Germany) and including UVIE (Austria); KU Leuven, CSL, IMEC (Belgium); CEA, LAM (France); MPIA (Germany); INAF-IFSI/OAA/OAP/OAT, LENS, SISSA (Italy); IAC (Spain). This development has been supported by the funding agencies BMVIT (Austria), ESA-PRODEX (Belgium), CEA/CNES (France), DLR (Germany), ASI/INAF (Italy), and CICYT/MCYT (Spain).
\end{acknowledgements}


\bibliographystyle{aa}
\bibliography{17109}

\begin{thebibliography}{100}
\expandafter\ifx\csname natexlab\endcsname\relax\def\natexlab#1{#1}\fi

\bibitem[{{Aikawa} {et~al.}(2008){Aikawa}, {Wakelam}, {Garrod}, \&
  {Herbst}}]{aikawa08a}
{Aikawa}, Y., {Wakelam}, V., {Garrod}, R.~T., \& {Herbst}, E. 2008, \apj, 674,
  984

\bibitem[{{Arce} \& {Sargent}(2006)}]{arce06a}
{Arce}, H.~G. \& {Sargent}, A.~I. 2006, \apj, 646, 1070

\bibitem[{{Arce} {et~al.}(2007){Arce}, {Shepherd}, {Gueth}, {Lee}, {Bachiller},
  {Rosen}, \& {Beuther}}]{arce07a}
{Arce}, H.~G., {Shepherd}, D., {Gueth}, F., {et~al.} 2007, Protostars \&
  Planets V, 245

\bibitem[{{Bachiller} \& {Tafalla}(1999)}]{bachiller99a}
{Bachiller}, R. \& {Tafalla}, M. 1999, in The Origin of Stars and Planetary
  Systems, ed. {C.~J.~Lada \& N.~D.~Kylafis} (Dordrecht: Kluwer), 227

\bibitem[{{Bakes} \& {Tielens}(1994)}]{bakes94a}
{Bakes}, E.~L.~O. \& {Tielens}, A.~G.~G.~M. 1994, \apj, 427, 822

\bibitem[{{Bergin} {et~al.}(1998){Bergin}, {Neufeld}, \& {Melnick}}]{bergin98a}
{Bergin}, E.~A., {Neufeld}, D.~A., \& {Melnick}, G.~J. 1998, \apj, 499, 777

\bibitem[{{Bisschop} {et~al.}(2006){Bisschop}, {Fraser}, {{\"O}berg}, {van
  Dishoeck}, \& {Schlemmer}}]{bisschop06a}
{Bisschop}, S.~E., {Fraser}, H.~J., {{\"O}berg}, K.~I., {van Dishoeck}, E.~F.,
  \& {Schlemmer}, S. 2006, \aap, 449, 1297

\bibitem[{{Bohlin} {et~al.}(1978){Bohlin}, {Savage}, \& {Drake}}]{bohlin78a}
{Bohlin}, R.~C., {Savage}, B.~D., \& {Drake}, J.~F. 1978, \apj, 224, 132

\bibitem[{{B{\"o}hm} {et~al.}(1993){B{\"o}hm}, {Noriega-Crespo}, \&
  {Solf}}]{bohm93a}
{B{\"o}hm}, K., {Noriega-Crespo}, A., \& {Solf}, J. 1993, \apj, 416, 647

\bibitem[{{Bontemps} {et~al.}(1996){Bontemps}, {Andre}, {Terebey}, \&
  {Cabrit}}]{bontemps96a}
{Bontemps}, S., {Andre}, P., {Terebey}, S., \& {Cabrit}, S. 1996, \aap, 311,
  858

\bibitem[{{Brinch} \& {Hogerheijde}(2010)}]{brinch10a}
{Brinch}, C. \& {Hogerheijde}, M.~R. 2010, \aap, 523, A25

\bibitem[{{Bruderer} {et~al.}(2010){Bruderer}, {Benz}, {St{\"a}uber}, \&
  {Doty}}]{bruderer10a}
{Bruderer}, S., {Benz}, A.~O., {St{\"a}uber}, P., \& {Doty}, S.~D. 2010, \apj,
  720, 1432

\bibitem[{{Bruderer} {et~al.}(2009){Bruderer}, {Doty}, \& {Benz}}]{bruderer09a}
{Bruderer}, S., {Doty}, S.~D., \& {Benz}, A.~O. 2009, \apjs, 183, 179

\bibitem[{{Ceccarelli} {et~al.}(1999){Ceccarelli}, {Caux}, {Loinard},
  {Castets}, {Tielens}, {Molinari}, {Liseau}, {Saraceno}, {Smith}, \&
  {White}}]{ceccarelli99a}
{Ceccarelli}, C., {Caux}, E., {Loinard}, L., {et~al.} 1999, \aap, 342, L21

\bibitem[{{Chavarr{\'\i}a} {et~al.}(2010){Chavarr{\'\i}a}, {Herpin}, {Jacq},
  {Braine}, {Bontemps}, {Baudry}, {Marseille}, {van der Tak}, {Pietropaoli},
  {Wyrowski}, {Shipman}, {Frieswijk}, {van Dishoeck}, {Cernicharo},
  {Bachiller}, {Benedettini}, {Benz}, {Bergin}, {Bjerkeli}, {Blake},
  {Bruderer}, {Caselli}, {Codella}, {Daniel}, {di Giorgio}, {Dominik}, {Doty},
  {Encrenaz}, {Fich}, {Fuente}, {Giannini}, {Goicoechea}, {de Graauw},
  {Hartogh}, {Helmich}, {Herczeg}, {Hogerheijde}, {Johnstone}, {J{\o}rgensen},
  {Kristensen}, {Larsson}, {Lis}, {Liseau}, {McCoey}, {Melnick}, {Nisini},
  {Olberg}, {Parise}, {Pearson}, {Plume}, {Risacher}, {Santiago-Garc{\'{\i}}a},
  {Saraceno}, {Stutzki}, {Szczerba}, {Tafalla}, {Tielens}, {van Kempen},
  {Visser}, {Wampfler}, {Willem}, \& {Y{\i}ld{\i}z}}]{chavarria10a}
{Chavarr{\'\i}a}, L., {Herpin}, F., {Jacq}, T., {et~al.} 2010, \aap, 521, L37

\bibitem[{{Curiel} {et~al.}(1995){Curiel}, {Raymond}, {Wolfire}, {Hartigan},
  {Morse}, {Schwartz}, \& {Nisenson}}]{curiel95a}
{Curiel}, S., {Raymond}, J.~C., {Wolfire}, M., {et~al.} 1995, \apj, 453, 322

\bibitem[{{Dalgarno}(2006)}]{dalgarno06a}
{Dalgarno}, A. 2006, Proc.\ Natl.\ Acad.\ Sci.\ USA, 103, 12269

\bibitem[{{de Graauw} {et~al.}(2010){de Graauw}, {Helmich}, {Phillips},
  {Stutzki}, {Caux}, {Whyborn}, {Dieleman}, {Roelfsema}, {Aarts}, {Assendorp},
  {Bachiller}, {Baechtold}, {Barcia}, {Beintema}, {Belitsky}, {Benz}, {Bieber},
  {Boogert}, {Borys}, {Bumble}, {Ca{\"i}s}, {Caris}, {Cerulli-Irelli},
  {Chattopadhyay}, {Cherednichenko}, {Ciechanowicz}, {Coeur-Joly}, {Comito},
  {Cros}, {de Jonge}, {de Lange}, {Delforges}, {Delorme}, {den Boggende},
  {Desbat}, {Diez-Gonz{\'a}lez}, {di Giorgio}, {Dubbeldam}, {Edwards},
  {Eggens}, {Erickson}, {Evers}, {Fich}, {Finn}, {Franke}, {Gaier}, {Gal},
  {Gao}, {Gallego}, {Gauffre}, {Gill}, {Glenz}, {Golstein}, {Goulooze},
  {Gunsing}, {G{\"u}sten}, {Hartogh}, {Hatch}, {Higgins}, {Honingh}, {Huisman},
  {Jackson}, {Jacobs}, {Jacobs}, {Jarchow}, {Javadi}, {Jellema}, {Justen},
  {Karpov}, {Kasemann}, {Kawamura}, {Keizer}, {Kester}, {Klapwijk}, {Klein},
  {Kollberg}, {Kooi}, {Kooiman}, {Kopf}, {Krause}, {Krieg}, {Kramer},
  {Kruizenga}, {Kuhn}, {Laauwen}, {Lai}, {Larsson}, {Leduc}, {Leinz}, {Lin},
  {Liseau}, {Liu}, {Loose}, {L{\'o}pez-Fernandez}, {Lord}, {Luinge}, {Marston},
  {Mart{\'{\i}}n-Pintado}, {Maestrini}, {Maiwald}, {McCoey}, {Mehdi}, {Megej},
  {Melchior}, {Meinsma}, {Merkel}, {Michalska}, {Monstein}, {Moratschke},
  {Morris}, {Muller}, {Murphy}, {Naber}, {Natale}, {Nowosielski}, {Nuzzolo},
  {Olberg}, {Olbrich}, {Orfei}, {Orleanski}, {Ossenkopf}, {Peacock}, {Pearson},
  {Peron}, {Phillip-May}, {Piazzo}, {Planesas}, {Rataj}, {Ravera}, {Risacher},
  {Salez}, {Samoska}, {Saraceno}, {Schieder}, {Schlecht}, {Schl{\"o}der},
  {Schm{\"u}lling}, {Schultz}, {Schuster}, {Siebertz}, {Smit}, {Szczerba},
  {Shipman}, {Steinmetz}, {Stern}, {Stokroos}, {Teipen}, {Teyssier}, {Tils},
  {Trappe}, {van Baaren}, {van Leeuwen}, {van de Stadt}, {Visser}, {Wildeman},
  {Wafelbakker}, {Ward}, {Wesselius}, {Wild}, {Wulff}, {Wunsch}, {Tielens},
  {Zaal}, {Zirath}, {Zmuidzinas}, \& {Zwart}}]{degraauw10a}
{de Graauw}, T., {Helmich}, F.~P., {Phillips}, T.~G., {et~al.} 2010, \aap, 518,
  L6

\bibitem[{{Di Francesco} {et~al.}(2008){Di Francesco}, {Johnstone}, {Kirk},
  {MacKenzie}, \& {Ledwosinska}}]{difrancesco08a}
{Di Francesco}, J., {Johnstone}, D., {Kirk}, H., {MacKenzie}, T., \&
  {Ledwosinska}, E. 2008, \apjs, 175, 277

\bibitem[{{Draine} \& {Bertoldi}(1996)}]{draine96a}
{Draine}, B.~T. \& {Bertoldi}, F. 1996, \apj, 468, 269

\bibitem[{{Dullemond} \& {Dominik}(2004)}]{dullemond04a}
{Dullemond}, C.~P. \& {Dominik}, C. 2004, \aap, 417, 159

\bibitem[{{Faure} {et~al.}(2007){Faure}, {Crimier}, {Ceccarelli}, {Valiron},
  {Wiesenfeld}, \& {Dubernet}}]{faure07a}
{Faure}, A., {Crimier}, N., {Ceccarelli}, C., {et~al.} 2007, \aap, 472, 1029

\bibitem[{{Fich} {et~al.}(2010){Fich}, {Johnstone}, {van Kempen}, {McCoey},
  {Fuente}, {Caselli}, {Kristensen}, {Plume}, {Cernicharo}, {Herczeg}, {van
  Dishoeck}, {Wampfler}, {Gaufre}, {Gill}, {Javadi}, {Justen}, {Laauwen},
  {Luinge}, {Ossenkopf}, {Pearson}, {Bachiller}, {Baudry}, {Benedettini},
  {Bergin}, {Benz}, {Bjerkeli}, {Blake}, {Bontemps}, {Braine}, {Bruderer},
  {Codella}, {Daniel}, {di Giorgio}, {Dominik}, {Doty}, {Encrenaz}, {Giannini},
  {Goicoechea}, {de Graauw}, {Helmich}, {Herpin}, {Hogerheijde}, {Jacq},
  {J{\o}rgensen}, {Larsson}, {Lis}, {Liseau}, {Marseille}, {Melnick}, {Nisini},
  {Olberg}, {Parise}, {Risacher}, {Santiago}, {Saraceno}, {Shipman}, {Tafalla},
  {van der Tak}, {Visser}, {Wyrowski}, \& {Y{\i}ld{\i}z}}]{fich10a}
{Fich}, M., {Johnstone}, D., {van Kempen}, T.~A., {et~al.} 2010, \aap, 518, L86

\bibitem[{{Flower} \& {Pineau des For{\^e}ts}(2003)}]{flower03a}
{Flower}, D.~R. \& {Pineau des For{\^e}ts}, G. 2003, \mnras, 343, 390

\bibitem[{{Flower} \& {Pineau Des For{\^e}ts}(2010)}]{flower10a}
{Flower}, D.~R. \& {Pineau Des For{\^e}ts}, G. 2010, \mnras, 406, 1745

\bibitem[{{Fraser} {et~al.}(2001){Fraser}, {Collings}, {McCoustra}, \&
  {Williams}}]{fraser01a}
{Fraser}, H.~J., {Collings}, M.~P., {McCoustra}, M.~R.~S., \& {Williams}, D.~A.
  2001, \mnras, 327, 1165

\bibitem[{{Froebrich}(2005)}]{froebrich05a}
{Froebrich}, D. 2005, \apjs, 156, 169

\bibitem[{{Giannini} {et~al.}(1999){Giannini}, {Lorenzetti}, {Tommasi},
  {Nisini}, {Benedettini}, {Pezzuto}, {Strafella}, {Barlow}, {Clegg}, {Cohen},
  {di Giorgio}, {Liseau}, {Molinari}, {Palla}, {Saraceno}, {Smith},
  {Spinoglio}, \& {White}}]{giannini99a}
{Giannini}, T., {Lorenzetti}, D., {Tommasi}, E., {et~al.} 1999, \aap, 346, 617

\bibitem[{{Giannini} {et~al.}(2001){Giannini}, {Nisini}, \&
  {Lorenzetti}}]{giannini01a}
{Giannini}, T., {Nisini}, B., \& {Lorenzetti}, D. 2001, \apj, 555, 40

\bibitem[{{Goldsmith} \& {Langer}(1999)}]{goldsmith99a}
{Goldsmith}, P.~F. \& {Langer}, W.~D. 1999, \apj, 517, 209

\bibitem[{{Graham} \& {Heyer}(1989)}]{graham89a}
{Graham}, J.~A. \& {Heyer}, M.~H. 1989, \pasp, 101, 573

\bibitem[{{Habart} {et~al.}(2010){Habart}, {Dartois}, {Abergel}, {Baluteau},
  {Naylor}, {Polehampton}, {Joblin}, {Ade}, {Anderson}, {Andr{\'e}}, {Arab},
  {Bernard}, {Blagrave}, {Bontemps}, {Boulanger}, {Cohen}, {Compiegne}, {Cox},
  {Davis}, {Emery}, {Fulton}, {Gry}, {Huang}, {Jones}, {Kirk}, {Lagache},
  {Lim}, {Madden}, {Makiwa}, {Martin}, {Miville-Desch{\^e}nes}, {Molinari},
  {Moseley}, {Motte}, {Okumura}, {Pinheiro Gon{\c c}alves}, {Rodon}, {Russeil},
  {Saraceno}, {Sidher}, {Spencer}, {Swinyard}, {Ward-Thompson}, {White}, \&
  {Zavagno}}]{habart10a}
{Habart}, E., {Dartois}, E., {Abergel}, A., {et~al.} 2010, \aap, 518, L116

\bibitem[{{Habing}(1968)}]{habing68a}
{Habing}, H.~J. 1968, \bain, 19, 421

\bibitem[{{Hirota} {et~al.}(2008){Hirota}, {Bushimata}, {Choi}, {Honma},
  {Imai}, {Iwadate}, {Jike}, {Kameya}, {Kamohara}, {Kan-Ya}, {Kawaguchi},
  {Kijima}, {Kobayashi}, {Kuji}, {Kurayama}, {Manabe}, {Miyaji}, {Nagayama},
  {Nakagawa}, {Oh}, {Omodaka}, {Oyama}, {Sakai}, {Sasao}, {Sato}, {Shibata},
  {Tamura}, \& {Yamashita}}]{hirota08a}
{Hirota}, T., {Bushimata}, T., {Choi}, Y.~K., {et~al.} 2008, \pasj, 60, 37

\bibitem[{{Hogerheijde} {et~al.}(1998){Hogerheijde}, {van Dishoeck}, {Blake},
  \& {van Langevelde}}]{hogerheijde98a}
{Hogerheijde}, M.~R., {van Dishoeck}, E.~F., {Blake}, G.~A., \& {van
  Langevelde}, H.~J. 1998, \apj, 502, 315

\bibitem[{{Hollenbach} \& {McKee}(1989)}]{hollenbach89a}
{Hollenbach}, D. \& {McKee}, C.~F. 1989, \apj, 342, 306

\bibitem[{{Hollenbach} {et~al.}(1991){Hollenbach}, {Takahashi}, \&
  {Tielens}}]{hollenbach91a}
{Hollenbach}, D.~J., {Takahashi}, T., \& {Tielens}, A.~G.~G.~M. 1991, \apj,
  377, 192

\bibitem[{{Hollenbach} \& {Tielens}(1997)}]{hollenbach97a}
{Hollenbach}, D.~J. \& {Tielens}, A.~G.~G.~M. 1997, \araa, 35, 179

\bibitem[{{Ingleby} {et~al.}(2011){Ingleby}, {Calvet}, {Hern{\'a}ndez},
  {Brice{\~n}o}, {Espaillat}, {Miller}, {Bergin}, \& {Hartmann}}]{ingleby11a}
{Ingleby}, L., {Calvet}, N., {Hern{\'a}ndez}, J., {et~al.} 2011, \aj, 141, 127

\bibitem[{{Ivezi{\'c}} {et~al.}(1999){Ivezi{\'c}}, {Nenkova}, \&
  {Elitzur}}]{ivezic99a}
{Ivezi{\'c}}, Z., {Nenkova}, M., \& {Elitzur}, M. 1999, {User Manual for DUSTY}
  (Univ.\ of Kentucky Internal Report, http://www.pa.uky.edu/$\sim$moshe/dusty)

\bibitem[{{Johnstone} {et~al.}(2010){Johnstone}, {Fich}, {McCoey}, {van
  Kempen}, {Fuente}, {Kristensen}, {Cernicharo}, {Caselli}, {Visser}, {Plume},
  {Herczeg}, {van Dishoeck}, {Wampfler}, {Bachiller}, {Baudry}, {Benedettini},
  {Bergin}, {Benz}, {Bjerkeli}, {Blake}, {Bontemps}, {Braine}, {Bruderer},
  {Codella}, {Daniel}, {di Giorgio}, {Dominik}, {Doty}, {Encrenaz}, {Giannini},
  {Goicoechea}, {de Graauw}, {Helmich}, {Herpin}, {Hogerheijde}, {Jacq},
  {J{\o}rgensen}, {Larsson}, {Lis}, {Liseau}, {Marseille}, {Melnick},
  {Neufeld}, {Nisini}, {Olberg}, {Parise}, {Pearson}, {Risacher},
  {Santiago-Garc{\'{\i}}a}, {Saraceno}, {Shipman}, {Tafalla}, {van der Tak},
  {Wyrowski}, {Y{\i}ld{\i}z}, {Caux}, {Honingh}, {Jellema}, {Schieder},
  {Teyssier}, \& {Whyborn}}]{johnstone10a}
{Johnstone}, D., {Fich}, M., {McCoey}, C., {et~al.} 2010, \aap, 521, L41

\bibitem[{{J{\o}rgensen} {et~al.}(2002){J{\o}rgensen}, {Sch{\"o}ier}, \& {van
  Dishoeck}}]{jorgensen02a}
{J{\o}rgensen}, J.~K., {Sch{\"o}ier}, F.~L., \& {van Dishoeck}, E.~F. 2002,
  \aap, 389, 908

\bibitem[{{J{\o}rgensen} {et~al.}(2005){J{\o}rgensen}, {Sch{\"o}ier}, \& {van
  Dishoeck}}]{jorgensen05c}
{J{\o}rgensen}, J.~K., {Sch{\"o}ier}, F.~L., \& {van Dishoeck}, E.~F. 2005,
  \aap, 435, 177

\bibitem[{{Kaufman} \& {Neufeld}(1996)}]{kaufman96a}
{Kaufman}, M.~J. \& {Neufeld}, D.~A. 1996, \apj, 456, 611

\bibitem[{{Kaufman} {et~al.}(1999){Kaufman}, {Wolfire}, {Hollenbach}, \&
  {Luhman}}]{kaufman99a}
{Kaufman}, M.~J., {Wolfire}, M.~G., {Hollenbach}, D.~J., \& {Luhman}, M.~L.
  1999, \apj, 527, 795 (K99)

\bibitem[{{Knee}(1992)}]{knee92a}
{Knee}, L.~B.~G. 1992, \aap, 259, 283

\bibitem[{{Kristensen} {et~al.}(2008){Kristensen}, {Ravkilde}, {Pineau des
  For{\^e}ts}, {Cabrit}, {Field}, {Gustafsson}, {Diana}, \&
  {Lemaire}}]{kristensen08a}
{Kristensen}, L.~E., {Ravkilde}, T.~L., {Pineau des For{\^e}ts}, G., {et~al.}
  2008, \aap, 477, 203

\bibitem[{{Kristensen} {et~al.}(2010){Kristensen}, {Visser}, {van Dishoeck},
  {Y{\i}ld{\i}z}, {Doty}, {Herczeg}, {Liu}, {Parise}, {J{\o}rgensen}, {van
  Kempen}, {Brinch}, {Wampfler}, {Bruderer}, {Benz}, {Hogerheijde}, {Deul},
  {Bachiller}, {Baudry}, {Benedettini}, {Bergin}, {Bjerkeli}, {Blake},
  {Bontemps}, {Braine}, {Caselli}, {Cernicharo}, {Codella}, {Daniel}, {de
  Graauw}, {di Giorgio}, {Dominik}, {Encrenaz}, {Fich}, {Fuente}, {Giannini},
  {Goicoechea}, {Helmich}, {Herpin}, {Jacq}, {Johnstone}, {Kaufman}, {Larsson},
  {Lis}, {Liseau}, {Marseille}, {McCoey}, {Melnick}, {Neufeld}, {Nisini},
  {Olberg}, {Pearson}, {Plume}, {Risacher}, {Santiago-Garc{\'{\i}}a},
  {Saraceno}, {Shipman}, {Tafalla}, {Tielens}, {van der Tak}, {Wyrowski},
  {Beintema}, {de Jonge}, {Dieleman}, {Ossenkopf}, {Roelfsema}, {Stutzki}, \&
  {Whyborn}}]{kristensen10b}
{Kristensen}, L.~E., {Visser}, R., {van Dishoeck}, E.~F., {et~al.} 2010, \aap,
  521, L30

\bibitem[{{Lada}(1985)}]{lada85a}
{Lada}, C.~J. 1985, \araa, 23, 267

\bibitem[{{Lesaffre} {et~al.}(2011){Lesaffre}, {Godard}, {Pineau des
  For{\^e}ts}, {Boulanger}, {Falgarone}, {Gerin}, \& {Guillard}}]{lesaffre11a}
{Lesaffre}, P., {Godard}, B., {Pineau des For{\^e}ts}, G., {et~al.} 2011, in
  IAU Symp.\ 280: The Molecular Universe, poster 2.052

\bibitem[{{Li} \& {Draine}(2001)}]{li01a}
{Li}, A. \& {Draine}, B.~T. 2001, \apj, 554, 778

\bibitem[{{Lloyd}(1982)}]{lloyd82a}
{Lloyd}, S.~P. 1982, IEEE T.\ Inform.\ Theory, 28, 129

\bibitem[{{Meeus} {et~al.}(2010){Meeus}, {Pinte}, {Woitke}, {Montesinos},
  {Mendigut{\'{\i}}a}, {Riviere-Marichalar}, {Eiroa}, {Mathews},
  {Vandenbussche}, {Howard}, {Roberge}, {Sandell}, {Duch{\^e}ne}, {M{\'e}nard},
  {Grady}, {Dent}, {Kamp}, {Augereau}, {Thi}, {Tilling}, {Alacid}, {Andrews},
  {Ardila}, {Aresu}, {Barrado}, {Brittain}, {Ciardi}, {Danchi}, {Fedele}, {de
  Gregorio-Monsalvo}, {Heras}, {Huelamo}, {Krivov}, {Lebreton}, {Liseau},
  {Martin-Zaidi}, {Mora}, {Morales-Calderon}, {Nomura}, {Pantin}, {Pascucci},
  {Phillips}, {Podio}, {Poelman}, {Ramsay}, {Riaz}, {Rice}, {Solano}, {Walker},
  {White}, {Williams}, \& {Wright}}]{meeus10a}
{Meeus}, G., {Pinte}, C., {Woitke}, P., {et~al.} 2010, \aap, 518, L124

\bibitem[{{Meijerink} {et~al.}(2007){Meijerink}, {Spaans}, \&
  {Israel}}]{meijerink07a}
{Meijerink}, R., {Spaans}, M., \& {Israel}, F.~P. 2007, \aap, 461, 793

\bibitem[{{Neufeld} {et~al.}(2009){Neufeld}, {Nisini}, {Giannini}, {Melnick},
  {Bergin}, {Yuan}, {Maret}, {Tolls}, {G{\"u}sten}, \& {Kaufman}}]{neufeld09a}
{Neufeld}, D.~A., {Nisini}, B., {Giannini}, T., {et~al.} 2009, \apj, 706, 170

\bibitem[{{Nishikawa} {et~al.}(2008){Nishikawa}, {Takami}, {Hayashi},
  {Wiseman}, \& {Pyo}}]{nishikawa08a}
{Nishikawa}, T., {Takami}, M., {Hayashi}, M., {Wiseman}, J., \& {Pyo}, T. 2008,
  \apj, 684, 1260

\bibitem[{{Nisini} {et~al.}(2002){Nisini}, {Giannini}, \&
  {Lorenzetti}}]{nisini02a}
{Nisini}, B., {Giannini}, T., \& {Lorenzetti}, D. 2002, \apj, 574, 246

\bibitem[{{{\"O}berg} {et~al.}(2007){{\"O}berg}, {Fuchs}, {Awad}, {Fraser},
  {Schlemmer}, {van Dishoeck}, \& {Linnartz}}]{oberg07a}
{{\"O}berg}, K.~I., {Fuchs}, G.~W., {Awad}, Z., {et~al.} 2007, \apjl, 662, L23

\bibitem[{{{\"O}berg} {et~al.}(2009){{\"O}berg}, {Linnartz}, {Visser}, \& {van
  Dishoeck}}]{oberg09a}
{{\"O}berg}, K.~I., {Linnartz}, H., {Visser}, R., \& {van Dishoeck}, E.~F.
  2009, \apj, 693, 1209

\bibitem[{{Ossenkopf} \& {Henning}(1994)}]{ossenkopf94a}
{Ossenkopf}, V. \& {Henning}, T. 1994, \aap, 291, 943

\bibitem[{{Ott}(2010)}]{ott10a}
{Ott}, S. 2010, in ASP Conf.\ Ser. 434: Astronomical Data Analysis Software and
  Systems XIX, ed. Y.~{Mizumoto}, K.-I. {Morita}, \& M.~{Ohishi} (San
  Francisco: ASP), 139

\bibitem[{{Pilbratt} {et~al.}(2010){Pilbratt}, {Riedinger}, {Passvogel},
  {Crone}, {Doyle}, {Gageur}, {Heras}, {Jewell}, {Metcalfe}, {Ott}, \&
  {Schmidt}}]{pilbratt10a}
{Pilbratt}, G.~L., {Riedinger}, J.~R., {Passvogel}, T., {et~al.} 2010, \aap,
  518, L1

\bibitem[{{Poglitsch} {et~al.}(2010){Poglitsch}, {Waelkens}, {Geis},
  {Feuchtgruber}, {Vandenbussche}, {Rodriguez}, {Krause}, {Renotte}, {van
  Hoof}, {Saraceno}, {Cepa}, {Kerschbaum}, {Agn{\`e}se}, {Ali}, {Altieri},
  {Andreani}, {Augueres}, {Balog}, {Barl}, {Bauer}, {Belbachir}, {Benedettini},
  {Billot}, {Boulade}, {Bischof}, {Blommaert}, {Callut}, {Cara}, {Cerulli},
  {Cesarsky}, {Contursi}, {Creten}, {De Meester}, {Doublier}, {Doumayrou},
  {Duband}, {Exter}, {Genzel}, {Gillis}, {Gr{\"o}zinger}, {Henning},
  {Herreros}, {Huygen}, {Inguscio}, {Jakob}, {Jamar}, {Jean}, {de Jong},
  {Katterloher}, {Kiss}, {Klaas}, {Lemke}, {Lutz}, {Madden}, {Marquet},
  {Martignac}, {Mazy}, {Merken}, {Montfort}, {Morbidelli}, {M{\"u}ller},
  {Nielbock}, {Okumura}, {Orfei}, {Ottensamer}, {Pezzuto}, {Popesso},
  {Putzeys}, {Regibo}, {Reveret}, {Royer}, {Sauvage}, {Schreiber}, {Stegmaier},
  {Schmitt}, {Schubert}, {Sturm}, {Thiel}, {Tofani}, {Vavrek}, {Wetzstein},
  {Wieprecht}, \& {Wiezorrek}}]{poglitsch10a}
{Poglitsch}, A., {Waelkens}, C., {Geis}, N., {et~al.} 2010, \aap, 518, L2

\bibitem[{{Raymond} {et~al.}(1997){Raymond}, {Blair}, \& {Long}}]{raymond97a}
{Raymond}, J.~C., {Blair}, W.~P., \& {Long}, K.~S. 1997, \apj, 489, 314

\bibitem[{{Robitaille} {et~al.}(2006){Robitaille}, {Whitney}, {Indebetouw},
  {Wood}, \& {Denzmore}}]{robitaille06a}
{Robitaille}, T.~P., {Whitney}, B.~A., {Indebetouw}, R., {Wood}, K., \&
  {Denzmore}, P. 2006, \apjs, 167, 256

\bibitem[{{R{\"o}llig} {et~al.}(2007){R{\"o}llig}, {Abel}, {Bell}, {Bensch},
  {Black}, {Ferland}, {Jonkheid}, {Kamp}, {Kaufman}, {Le Bourlot}, {Le Petit},
  {Meijerink}, {Morata}, {Ossenkopf}, {Roueff}, {Shaw}, {Spaans}, {Sternberg},
  {Stutzki}, {Thi}, {van Dishoeck}, {van Hoof}, {Viti}, \&
  {Wolfire}}]{rollig07a}
{R{\"o}llig}, M., {Abel}, N.~P., {Bell}, T., {et~al.} 2007, \aap, 467, 187

\bibitem[{{Sandell} {et~al.}(1994){Sandell}, {Knee}, {Aspin}, {Robson}, \&
  {Russell}}]{sandell94a}
{Sandell}, G., {Knee}, L.~B.~G., {Aspin}, C., {Robson}, I.~E., \& {Russell},
  A.~P.~G. 1994, \aap, 285, L1

\bibitem[{{Santiago-Garc{\'{\i}}a} {et~al.}(2009){Santiago-Garc{\'{\i}}a},
  {Tafalla}, {Johnstone}, \& {Bachiller}}]{santiago09a}
{Santiago-Garc{\'{\i}}a}, J., {Tafalla}, M., {Johnstone}, D., \& {Bachiller},
  R. 2009, \aap, 495, 169

\bibitem[{{Saucedo} {et~al.}(2003){Saucedo}, {Calvet}, {Hartmann}, \&
  {Raymond}}]{saucedo03a}
{Saucedo}, J., {Calvet}, N., {Hartmann}, L., \& {Raymond}, J. 2003, \apj, 591,
  275

\bibitem[{{Sch{\"o}ier} {et~al.}(2002){Sch{\"o}ier}, {J{\o}rgensen}, {van
  Dishoeck}, \& {Blake}}]{schoier02a}
{Sch{\"o}ier}, F.~L., {J{\o}rgensen}, J.~K., {van Dishoeck}, E.~F., \& {Blake},
  G.~A. 2002, \aap, 390, 1001

\bibitem[{{Sch{\"o}ier} {et~al.}(2004){Sch{\"o}ier}, {J{\o}rgensen}, {van
  Dishoeck}, \& {Blake}}]{schoier04a}
{Sch{\"o}ier}, F.~L., {J{\o}rgensen}, J.~K., {van Dishoeck}, E.~F., \& {Blake},
  G.~A. 2004, \aap, 418, 185

\bibitem[{{Sch{\"o}ier} {et~al.}(2005){Sch{\"o}ier}, {van der Tak}, {van
  Dishoeck}, \& {Black}}]{schoier05a}
{Sch{\"o}ier}, F.~L., {van der Tak}, F.~F.~S., {van Dishoeck}, E.~F., \&
  {Black}, J.~H. 2005, \aap, 432, 369

\bibitem[{{Schuster} {et~al.}(1993){Schuster}, {Harris}, {Anderson}, \&
  {Russell}}]{schuster93a}
{Schuster}, K.~F., {Harris}, A.~I., {Anderson}, N., \& {Russell}, A.~P.~G.
  1993, \apjl, 412, L67

\bibitem[{{Shang} {et~al.}(2006){Shang}, {Allen}, {Li}, {Liu}, {Chou}, \&
  {Anderson}}]{shang06a}
{Shang}, H., {Allen}, A., {Li}, Z., {et~al.} 2006, \apj, 649, 845

\bibitem[{{Shen} {et~al.}(2004){Shen}, {Greenberg}, {Schutte}, \& {van
  Dishoeck}}]{shen04a}
{Shen}, C.~J., {Greenberg}, J.~M., {Schutte}, W.~A., \& {van Dishoeck}, E.~F.
  2004, \aap, 415, 203

\bibitem[{{Shu}(1977)}]{shu77a}
{Shu}, F.~H. 1977, \apj, 214, 488

\bibitem[{{Shu} {et~al.}(1987){Shu}, {Adams}, \& {Lizano}}]{shu87a}
{Shu}, F.~H., {Adams}, F.~C., \& {Lizano}, S. 1987, \araa, 25, 23

\bibitem[{{Spaans} {et~al.}(1995){Spaans}, {Hogerheijde}, {Mundy}, \& {van
  Dishoeck}}]{spaans95a}
{Spaans}, M., {Hogerheijde}, M.~R., {Mundy}, L.~G., \& {van Dishoeck}, E.~F.
  1995, \apjl, 455, L167

\bibitem[{{Sturm} {et~al.}(2010){Sturm}, {Bouwman}, {Henning}, {Evans}, {Acke},
  {Mulders}, {Waters}, {van Dishoeck}, {Meeus}, {Green}, {Augereau},
  {Olofsson}, {Salyk}, {Najita}, {Herczeg}, {van Kempen}, {Kristensen},
  {Dominik}, {Carr}, {Waelkens}, {Bergin}, {Blake}, {Brown}, {Chen}, {Cieza},
  {Dunham}, {Glassgold}, {G{\"u}del}, {Harvey}, {Hogerheijde}, {Jaffe},
  {J{\o}rgensen}, {Kim}, {Knez}, {Lacy}, {Lee}, {Maret}, {Meijerink},
  {Mer{\'{\i}}n}, {Mundy}, {Pontoppidan}, {Visser}, \&
  {Y{\i}ld{\i}z}}]{sturm10a}
{Sturm}, B., {Bouwman}, J., {Henning}, T., {et~al.} 2010, \aap, 518, L129

\bibitem[{{Tielens} \& {Hollenbach}(1985)}]{tielens85a}
{Tielens}, A.~G.~G.~M. \& {Hollenbach}, D. 1985, \apj, 291, 722

\bibitem[{{Tobin} {et~al.}(2008){Tobin}, {Hartmann}, {Calvet}, \&
  {D'Alessio}}]{tobin08a}
{Tobin}, J.~J., {Hartmann}, L., {Calvet}, N., \& {D'Alessio}, P. 2008, \apj,
  679, 1364

\bibitem[{{van der Werf} {et~al.}(2010){van der Werf}, {Isaak}, {Meijerink},
  {Spaans}, {Rykala}, {Fulton}, {Loenen}, {Walter}, {Wei{\ss}}, {Armus},
  {Fischer}, {Israel}, {Harris}, {Veilleux}, {Henkel}, {Savini}, {Lord},
  {Smith}, {Gonz{\'a}lez-Alfonso}, {Naylor}, {Aalto}, {Charmandaris}, {Dasyra},
  {Evans}, {Gao}, {Greve}, {G{\"u}sten}, {Kramer}, {Mart{\'{\i}}n-Pintado},
  {Mazzarella}, {Papadopoulos}, {Sanders}, {Spinoglio}, {Stacey}, {Vlahakis},
  {Wiedner}, \& {Xilouris}}]{vanderwerf10a}
{van der Werf}, P.~P., {Isaak}, K.~G., {Meijerink}, R., {et~al.} 2010, \aap,
  518, L42

\bibitem[{{van Dishoeck} {et~al.}(2006){van Dishoeck}, {Jonkheid}, \& {van
  Hemert}}]{vandishoeck06a}
{van Dishoeck}, E.~F., {Jonkheid}, B., \& {van Hemert}, M.~C. 2006, Faraday\
  Discuss., 133, 231

\bibitem[{{van Dishoeck} {et~al.}(2011){van Dishoeck}, {Kristensen}, {Benz},
  {Bergin}, {Caselli}, {Cernicharo}, {Herpin}, {Hogerheijde}, {Johnstone},
  {Liseau}, {Nisini}, {Shipman}, {Tafalla}, {van der Tak}, {Wyrowski},
  {Aikawa}, {Bachiller}, {Baudry}, {Benedettini}, {Bjerkeli}, {Blake},
  {Bontemps}, {Braine}, {Brinch}, {Bruderer}, {Chavarr{\'{\i}}a}, {Codella},
  {Daniel}, {de Graauw}, {Deul}, {di Giorgio}, {Dominik}, {Doty}, {Dubernet},
  {Encrenaz}, {Feuchtgruber}, {Fich}, {Frieswijk}, {Fuente}, {Giannini},
  {Goicoechea}, {Helmich}, {Herczeg}, {Jacq}, {J{\o}rgensen}, {Karska},
  {Kaufman}, {Keto}, {Larsson}, {Lefloch}, {Lis}, {Marseille}, {McCoey},
  {Melnick}, {Neufeld}, {Olberg}, {Pagani}, {Pani{\'c}}, {Parise}, {Pearson},
  {Plume}, {Risacher}, {Salter}, {Santiago-Garc{\'{\i}}a}, {Saraceno},
  {St{\"a}uber}, {van Kempen}, {Visser}, {Viti}, {Walmsley}, {Wampfler}, \&
  {Y{\i}ld{\i}z}}]{vandishoeck11a}
{van Dishoeck}, E.~F., {Kristensen}, L.~E., {Benz}, A.~O., {et~al.} 2011,
  \pasp, 123, 138

\bibitem[{{van Kempen} {et~al.}(2010{\natexlab{a}}){van Kempen}, {Green},
  {Evans}, {van Dishoeck}, {Kristensen}, {Herczeg}, {Mer{\'{\i}}n}, {Lee},
  {J{\o}rgensen}, {Bouwman}, {Acke}, {Adamkovics}, {Augereau}, {Bergin},
  {Blake}, {Brown}, {Carr}, {Chen}, {Cieza}, {Dominik}, {Dullemond}, {Dunham},
  {Glassgold}, {G{\"u}del}, {Harvey}, {Henning}, {Hogerheijde}, {Jaffe}, {Kim},
  {Knez}, {Lacy}, {Maret}, {Meeus}, {Meijerink}, {Mulders}, {Mundy}, {Najita},
  {Olofsson}, {Pontoppidan}, {Salyk}, {Sturm}, {Visser}, {Waters}, {Waelkens},
  \& {Y{\i}ld{\i}z}}]{vankempen10b}
{van Kempen}, T.~A., {Green}, J.~D., {Evans}, N.~J., {et~al.}
  2010{\natexlab{a}}, \aap, 518, L128

\bibitem[{{van Kempen} {et~al.}(2006){van Kempen}, {Hogerheijde}, {van
  Dishoeck}, {G{\"u}sten}, {Schilke}, \& {Nyman}}]{vankempen06a}
{van Kempen}, T.~A., {Hogerheijde}, M.~R., {van Dishoeck}, E.~F., {et~al.}
  2006, \aap, 454, L75

\bibitem[{{van Kempen} {et~al.}(2010{\natexlab{b}}){van Kempen}, {Kristensen},
  {Herczeg}, {Visser}, {van Dishoeck}, {Wampfler}, {Bruderer}, {Benz}, {Doty},
  {Brinch}, {Hogerheijde}, {J{\o}rgensen}, {Tafalla}, {Neufeld}, {Bachiller},
  {Baudry}, {Benedettini}, {Bergin}, {Bjerkeli}, {Blake}, {Bontemps}, {Braine},
  {Caselli}, {Cernicharo}, {Codella}, {Daniel}, {di Giorgio}, {Dominik},
  {Encrenaz}, {Fich}, {Fuente}, {Giannini}, {Goicoechea}, {de Graauw},
  {Helmich}, {Herpin}, {Jacq}, {Johnstone}, {Kaufman}, {Larsson}, {Lis},
  {Liseau}, {Marseille}, {McCoey}, {Melnick}, {Nisini}, {Olberg}, {Parise},
  {Pearson}, {Plume}, {Risacher}, {Santiago-Garc{\'{\i}}a}, {Saraceno},
  {Shipman}, {van der Tak}, {Wyrowski}, {Y{\i}ld{\i}z}, {Ciechanowicz},
  {Dubbeldam}, {Glenz}, {Huisman}, {Lin}, {Morris}, {Murphy}, \&
  {Trappe}}]{vankempen10a}
{van Kempen}, T.~A., {Kristensen}, L.~E., {Herczeg}, G.~J., {et~al.}
  2010{\natexlab{b}}, \aap, 518, L121

\bibitem[{{van Kempen} {et~al.}(2009){van Kempen}, {van Dishoeck},
  {G{\"u}sten}, {Kristensen}, {Schilke}, {Hogerheijde}, {Boland}, {Nefs},
  {Menten}, {Baryshev}, \& {Wyrowski}}]{vankempen09b}
{van Kempen}, T.~A., {van Dishoeck}, E.~F., {G{\"u}sten}, R., {et~al.} 2009,
  \aap, 501, 633

\bibitem[{{van Zadelhoff} {et~al.}(2003){van Zadelhoff}, {Aikawa},
  {Hogerheijde}, \& {van Dishoeck}}]{vanzadelhoff03a}
{van Zadelhoff}, G.-J., {Aikawa}, Y., {Hogerheijde}, M.~R., \& {van Dishoeck},
  E.~F. 2003, \aap, 397, 789

\bibitem[{{Velusamy} {et~al.}(2007){Velusamy}, {Langer}, \&
  {Marsh}}]{velusamy07a}
{Velusamy}, T., {Langer}, W.~D., \& {Marsh}, K.~A. 2007, \apjl, 668, L159

\bibitem[{{Visser} {et~al.}(2009){Visser}, {van Dishoeck}, \&
  {Black}}]{visser09b}
{Visser}, R., {van Dishoeck}, E.~F., \& {Black}, J.~H. 2009, \aap, 503, 323

\bibitem[{{Wagner} \& {Graff}(1987)}]{wagner87a}
{Wagner}, A.~F. \& {Graff}, M.~M. 1987, \apj, 317, 423

\bibitem[{{Walter} {et~al.}(2003){Walter}, {Herczeg}, {Brown}, {Ardila},
  {Gahm}, {Johns-Krull}, {Lissauer}, {Simon}, \& {Valenti}}]{walter03a}
{Walter}, F.~M., {Herczeg}, G., {Brown}, A., {et~al.} 2003, \aj, 126, 3076

\bibitem[{{Wampfler} {et~al.}(2010){Wampfler}, {Herczeg}, {Bruderer}, {Benz},
  {van Dishoeck}, {Kristensen}, {Visser}, {Doty}, {Melchior}, {van Kempen},
  {Y{\i}ld{\i}z}, {Dedes}, {Goicoechea}, {Baudry}, {Melnick}, {Bachiller},
  {Benedettini}, {Bergin}, {Bjerkeli}, {Blake}, {Bontemps}, {Braine},
  {Caselli}, {Cernicharo}, {Codella}, {Daniel}, {di Giorgio}, {Dominik},
  {Encrenaz}, {Fich}, {Fuente}, {Giannini}, {de Graauw}, {Helmich}, {Herpin},
  {Hogerheijde}, {Jacq}, {Johnstone}, {J{\o}rgensen}, {Larsson}, {Lis},
  {Liseau}, {Marseille}, {McCoey}, {Neufeld}, {Nisini}, {Olberg}, {Parise},
  {Pearson}, {Plume}, {Risacher}, {Santiago-Garc{\'{\i}}a}, {Saraceno},
  {Shipman}, {Tafalla}, {van der Tak}, {Wyrowski}, {Roelfsema}, {Jellema},
  {Dieleman}, {Caux}, \& {Stutzki}}]{wampfler10a}
{Wampfler}, S.~F., {Herczeg}, G.~J., {Bruderer}, S., {et~al.} 2010, \aap, 521,
  L36

\bibitem[{{Whitney} {et~al.}(2003){Whitney}, {Wood}, {Bjorkman}, \&
  {Cohen}}]{whitney03b}
{Whitney}, B.~A., {Wood}, K., {Bjorkman}, J.~E., \& {Cohen}, M. 2003, \apj,
  598, 1079

\bibitem[{{Whittet} {et~al.}(1997){Whittet}, {Prusti}, {Franco}, {Gerakines},
  {Kilkenny}, {Larson}, \& {Wesselius}}]{whittet97a}
{Whittet}, D.~C.~B., {Prusti}, T., {Franco}, G.~A.~P., {et~al.} 1997, \aap,
  327, 1194

\bibitem[{{Woodall} {et~al.}(2007){Woodall}, {Ag{\'u}ndez}, {Markwick-Kemper},
  \& {Millar}}]{woodall07a}
{Woodall}, J., {Ag{\'u}ndez}, M., {Markwick-Kemper}, A.~J., \& {Millar}, T.~J.
  2007, \aap, 466, 1197

\bibitem[{{Yang} {et~al.}(2010){Yang}, {Stancil}, {Balakrishnan}, \&
  {Forrey}}]{yang10a}
{Yang}, B., {Stancil}, P.~C., {Balakrishnan}, N., \& {Forrey}, R.~C. 2010,
  \apj, 718, 1062

\bibitem[{{Y{\i}ld{\i}z} {et~al.}(2010){Y{\i}ld{\i}z}, {van Dishoeck},
  {Kristensen}, {Visser}, {J{\o}rgensen}, {Herczeg}, {van Kempen},
  {Hogerheijde}, {Doty}, {Benz}, {Bruderer}, {Wampfler}, {Deul}, {Bachiller},
  {Baudry}, {Benedettini}, {Bergin}, {Bjerkeli}, {Blake}, {Bontemps}, {Braine},
  {Caselli}, {Cernicharo}, {Codella}, {Daniel}, {di Giorgio}, {Dominik},
  {Encrenaz}, {Fich}, {Fuente}, {Giannini}, {Goicoechea}, {de Graauw},
  {Helmich}, {Herpin}, {Jacq}, {Johnstone}, {Larsson}, {Lis}, {Liseau}, {Liu},
  {Marseille}, {McCoey}, {Melnick}, {Neufeld}, {Nisini}, {Olberg}, {Parise},
  {Pearson}, {Plume}, {Risacher}, {Santiago-Garc{\'{\i}}a}, {Saraceno},
  {Shipman}, {Tafalla}, {Tielens}, {van der Tak}, {Wyrowski}, {Dieleman},
  {Jellema}, {Ossenkopf}, {Schieder}, \& {Stutzki}}]{yildiz10a}
{Y{\i}ld{\i}z}, U.~A., {van Dishoeck}, E.~F., {Kristensen}, L.~E., {et~al.}
  2010, \aap, 521, L40

\bibitem[{{Young} {et~al.}(2004){Young}, {Lee}, {Evans}, {Goldsmith}, \&
  {Doty}}]{young04a}
{Young}, K.~E., {Lee}, J., {Evans}, II, N.~J., {Goldsmith}, P.~F., \& {Doty},
  S.~D. 2004, \apj, 614, 252

\end{thebibliography}

\end{document}